\documentclass[11pt]{article}
\usepackage{amsmath,euscript,amssymb}
\usepackage[T2A]{fontenc}
\usepackage[cp866]{inputenc}
\usepackage[dvips]{graphicx}
\def\v1{\vspace{1cm}}
\def\be{\begin{equation}}
\def\ee{\end{equation}}
\def\bc{\begin{center}}
\def\ec{\end{center}}

\def\vh{\varphi}

\newcommand{\bea}{\begin{eqnarray}}
\newcommand{\eea}{\end{eqnarray}}

\begin{document}
\title{
Hamiltonian   General Relativity  in  Finite Space\\
 and Cosmological Potential Perturbations
 }
 \author{ B.M. Barbashov${}^{1}$,  V.N. Pervushin${}^{1}$, A.F. Zakharov${}^{1,2,3,4}$, and
V.A. Zinchuk${}^{1}$,
 \\[0.3cm]
{\normalsize\it $^1$ Joint Institute for Nuclear Research},\\
 {\normalsize\it 141980, Dubna, Russia} \\
 {\normalsize\it $^2$ National Astronomical Observatories of Chinese Academy
of Sciences},\\
 {\normalsize\it  Beijing 100012, China}\\
{\normalsize\it $^3$ Institute of Theoretical and Experimental
Physics, 25,
117259, Moscow}\\
{\normalsize\it $^4$ Astro Space Center of Lebedev Physics Institute of RAS, Moscow}\\
  }

\date{\empty}
\maketitle
\medskip

\begin{abstract}
 {\noindent
  The Hamiltonian formulation of general relativity (GR)
  is  considered in  finite space-time and a specific
  reference frame given by the diffeo-invariant components of
  the Fock simplex in terms of the Dirac -- ADM variables.

   The evolution parameter and energy   invariant
  with respect to the time-coordinate transformations are constructed
  by the separation of
 the cosmological scale factor $a(x^0)$ and its identification  with the
 spatial averaging of the metric determinant, so that the
 dimension of the kinemetric  group of diffeomorphisms coincides with
 the  dimension of a set of variables whose velocities are removed
 by the Gauss-type constraints in accordance with the second N\"other theorem.
   This  coincidence
  allows us to solve the energy constraint,
 fulfil Dirac's Hamiltonian reduction,  and to
describe the potential perturbations in terms of  the Lichnerowicz
scale-invariant variables distinguished by the absence of the time
derivatives of the spatial metric determinant. It was shown that
the Hamiltonian version of the cosmological perturbation
 theory acquires attributes of the theory of superfluid
 liquid, and it leads to a generalization of the Schwarzschild
 solution.

 The astrophysical application of this approach to GR is
 considered under supposition that the Dirac -- ADM Hamiltonian
 frame is identified with that of the Cosmic Microwave Background
 radiation  distinguished by its dipole component in the frame
 of an Earth observer.

}
\end{abstract}
%\end{titlepage}

%\vspace{-17cm}
%
%\hfill {\small Is devoted to the 90th anniversary of David
%Hilbert's talk}
%
%\hfill   {\small {\it  ``Die Grundlangen der Physik''} on 20
%November 1915 in G\"ottingen.}
%
%\vspace{18cm}

%\vspace{2cm}
%
% \centerline{\bf Invited Talk at Int. Conference ``New
%trends in High-Energy Physics''}
%
% \centerline{\bf Yalta, Crimea,
%Ukraine, September 10 - 17, 2005}
%
%\vspace{1cm}

 %\centerline{ To be published in Proceedings of
% Int. Conference}
%
% \centerline{ ``New
%trends in High-Energy Physics''}
%
%\centerline{ Yalta, Crimea, Ukraine, September 10 - 17, 2005.}
%
% \centerline{  The NATO ARW ``Nuclear Science and Safety in
%Europe''}

\newpage

\tableofcontents

\section{Introduction}

 In the year of  celebration of the 90th
 anniversary of general relativity (GR) \cite{einsh,H}
 one can  distinguish two treatments of general coordinate
 transformations  i.) as
   generalization of the {\it frame transformations}
   and ii.) as  {\it diffeomorphisms} of the GR action
 and a {\it geometric
 interval}.
  There is
  an essential difference between the frame group
 of the Lorentz -- Poincar\'e-type  \cite{poi}
  leading to a set of
 initial data
 and the {\it diffeomorphism  group} of general coordinate transformations
  restricting these initial data by {\it constraints}. This difference
  was revealed by two  N\"other theorems  \cite{Noter}\footnote{One of
  these theorems (the second) was  formulated by Hilbert in his famous paper \cite{H}
   (see also its revised version  \cite{H24}). We should like to thank
   V.V. Nesterenko who draw our attention to this fact \cite{d1986}.}.
  It became evident in the light of the formulation
  of GR in terms of
 the Fock simplex \cite{fock29} defined as a diffeo-invariant Lorentz
 vector that helps us  to separate  diffeomorphisms from transformations of
reference frames.

  Just this separation  of diffeomorphisms from the frame transformations
  is  a starting position of this paper
  devoted to the consideration of GR
  in  finite space-time in
  a specific reference frame given by the diffeo-invariant components of
  the Fock simplex in terms of the Dirac -- ADM variables \cite{dir}
  widely used for
 the Hamiltonian formulation of GR. Fixing the ``Hamiltonian frame'' one can
 determine the group of
 diffeomorphisms of this frame.

 Another starting position of the paper is the kinemetric group of
 diffeomorphisms of this ``Hamiltonian frame''
 established in \cite{vlad}. This kinemetric group contains
  global parametrizations of the coordinate time and  local
  transformations of three spatial coordinates, and it
   requires
  to distinguish
 a set of variables with the same dimension,
 velocities (or momenta) of which
 are removed by the Gauss-type constraints
 from the phase space of diffeo-invariant physical
 variables in  accordance
  with the second N\"other theorem.

A similar Hilbert-type \cite{H,H24} {\it geometro}-dynamic
  formulation of  special relativity (SR) \cite{WDW,pp,bpp}
  with reparameterizations
  of the coordinate evolution parameter shows that
 the energy constraint fixes
 a velocity of one of the dynamic variables
 that becomes a diffeo-invariant dynamic evolution parameter.
In particular, in SR such a
 dynamic evolution parameter is well known, it is the
 fourth component of the Minkowskian space-time
 coordinate vector.

 In order to realize a similar construction of
  the
  Dirac -- ADM Hamiltonian GR \cite{dir},
 one should  point out  in GR a homogeneous variable that can be
  a diffeo-invariant evolution parameter in the field space of
  events \cite{WDW}
   in accordance
  with the kinemetric diffeomorphism group of GR in the ``Hamiltonian frame''
  \cite{vlad}.
 The cosmological evolution is the irrefutable observational
 argument in favor of
 existence of such a homogeneous variable  considered in GR
  as the cosmological scale factor.

 The separation of the cosmological scale factor from metrics in GR
 is well-known as the cosmological perturbation theory proposed by
 Lifshits \cite{Lif,kodama} and applied as the basic tools
 for analysis of modern observational data in astrophysics and cosmology \cite{bard}.
 However, as it was shown in \cite{242,bpzz},
  the Hamiltonian analysis of the standard Lifshits
 perturbation theory  reveals that  perturbation of
 the spatial metric determinant should be split  from the cosmological
 scale factor  by
 the projection operator onto nonhomogeneous class functions,
 otherwise the determinant perturbations
    contain one more variable equivalent to the scale factor
   that is the
 obstruction to the Hamiltonian approach.

 Thus, in the Dirac approach to GR
 this homogeneous evolution parameter is not split; whereas
  the standard cosmological perturbation theory contains two
  such-type variables; so that in both the cases the
  dimension of the Hamiltonian constraints does not coincide with
  the dimension of the diffeomorphism group, which
  contradicts  the second N\"other theorem.

 In the present paper, in order to restore the number of variables
 of the initial GR,
 a homogeneous evolution parameter as
  a cosmological scale factor $a(x^0)$ is identified with
 the spatial averaging the metric determinant in the
Dirac - ADM Hamiltonian reference frame with a finite space-time.

  We show that this formulation of GR
  can solve the
 ``energy-time'' problem in both classical and quantum GR, and the latter
 has some attributes of the
 theory of superfluidity. In particular, there are spatial
 determinant free (i.e scale-invariant)
  variables introduced by Lichnerowicz \cite{L,Y}, in terms of
  which the
 classical
 GR contains only the Landau-type ``friction-free'' dynamics
 independent
  of the velocity-velocity interaction  \cite{Lan}. It leads to
  resolution of the energy constraint with respect to the
  cosmological scale momentum so that its positive and negative values become
  the generators of evolution of all scale-invariant field
  variables. The negative energy problem is solved by the primary
  and secondary quantization that leads to
  London's unique wave function \cite{Lon}
  and   Bogoliubov's squeezed condensate  \cite{B,origin}, respectively.
   All these attributes
  are accompanied by a set of physical consequences that can be
  understood as  quantum effects.

  The diffeo-invariant Hamiltonian version of cosmological perturbation theory
is applied in order to consider  topical
 problems of modern cosmology by
 the low-energy
 decomposition of the reduced action in terms of the Lichnerowicz variables
 that identify conformal quantities with observables. It means
 that   the Hubble law is explained
 by
 the evolution of masses, so that  the Dark Energy chosen as  the homogeneous
 free scalar field \cite{039,039a,Danilo} is in agreement with
 primordial nucleosynthesis \cite{three} and the last Supernova
 data \cite{snov,SN}.
 In this approach to GR and Standard
Model (SM)  all matter (including CMB radiation) appears as a final
decay product of primordial vector W-, Z- bosons cosmologically
created from the vacuum when their Compton length coincides with the
universe horizon \cite{114:a,vin}.
 The equations describing longitudinal
 vector bosons
 in SM, in this case, become close to the
  equations of the
 Lifshits perturbation theory which are
 used in the classical inflationary model for
 description of the ``power primordial spectrum'' of the CMB radiation \cite{cmb}.
 This means that the considered reference frame is
 identified with the CMB radiation one.

\section{The Dirac -- ADM Hamiltonian approach in terms of a simplex}

\subsection{GR in terms of Fock's  simplex of reference}

     GR
    is given by two fundamental quantities:
    the  {\it``dynamic''} action
 \be\label{gr}
 S=\int d^4x\sqrt{-g}\left[-\frac{\vh_0^2}{6}R(g)
 +{\cal L}_{(\rm M)}\right]\equiv S_{\rm GR}+{S}_{\rm M},
 \ee
 where  $\varphi^2_0=\dfrac{3}{8\pi}M^2_{\rm Planck}=
 \dfrac{3}{8\pi G}$, $G$ is the Newton
constant in the units $\hbar=c=1$,
 ${\cal L}_{(\rm M)}$  is the  Lagrangian of the matter field,
     and a
 {\it``geometric interval''}
\be \label{ds}
 g_{\mu\nu}dx^\mu dx^\nu\equiv\omega_{(\alpha)}\omega_{(\alpha)}=
 \omega_{(0)}\omega_{(0)}-
 \omega_{(1)}\omega_{(1)}-\omega_{(2)}\omega_{(2)}-\omega_{(3)}\omega_{(3)},
 \ee
 where $\omega_{(\alpha)}$  are linear differential forms as components of an orthogonal
 simplex of reference \cite{fock29}.

 GR
 in terms of Fock's  simplex  %(\ref{gr}), (\ref{ds})
  contains  two principles of
 relativity: the {\it``geometric''} --- general coordinate transformations
\be \label{1zel}
 x^{\mu} \rightarrow  \tilde x^{\mu}=\tilde
 x^{\mu}(x^0,x^{1},x^{2},x^{3}),~~~~~~~~~
 \omega_{(\alpha)}(x^{\mu})~\to ~\omega_{(\alpha)}(\tilde x^{\mu})=
 \omega_{(\alpha)}(x^{\mu})
 \ee
 and the set of  transformations  of a reference frame
 identified with the Lorentz
 transformations of an orthogonal  simplex of reference
 \be \label{2zel}
{\omega}_{(\alpha)}~\to ~
\overline{\omega}_{(\alpha)}=L_{(\alpha)(\beta)}{\omega}_{(\beta)}.
\ee
 The invariance of the action with respect to
 frame transformations means that there are
 integrals of motion (the first N\"other theorem \cite{Noter}); whereas the invariance
  of the action with respect to
 diffeomorphisms  means that a part of
 degrees of freedom corresponds to pure gauge
  non-dynamical variables  and diffeo-invariant potentials defined
  by constraints.

\subsection{Diffeo-invariant variables and potentials}

  According to the second N\"other theorem \cite{Noter},
 in any theory invariant
 with respect to diffeomorphisms there are constraints
 of velocities (momenta) that remove part of variables as pure
 gauge ones. These constraints
 are established  in the specific reference frame distinguished
 by the Lorentz vector $l_\mu=(1,0,0,0)$ \cite{cj}.

 In particular,  in   QED  in the Minkowskian space-time
 given by equations
 $\partial_\mu F^{\mu \nu}=J^\nu$,
 the invariance of the theory with respect to diffeomorphisms
 well known as gauge transformations $A_\mu \to \widetilde{A}_\mu=
 A_\mu+\partial_\mu\lambda$, $\Psi \to \widetilde{\Psi}=e^{ie\lambda}\Psi$
  allows us to remove the pure gauge longitudinal component,
 if $\lambda=\lambda^T$ is chosen so that it satisfies the
 equation
 $\triangle\lambda^T=\partial_kA_k$.  After this transformation
 the zero component of the equations $\partial_\mu F^{\mu 0}=J^0$
 known as the Gauss constraint  takes the form
 $\triangle A^T_0=J^T_0$ (see \cite{cj}).
 Thus, all four components of the vector field
 $A_\mu=(A_0,A_k)$
 can be  split into the pure ``gauge'' longitudinal component
 $\partial_k A_k$, a diffeo-invariant
 potential $\triangle A^T_0=\triangle A_0-\partial_k \partial_0
 A_k$,
 and two diffeo-invariant transverse dynamic variables $A_i^{T}$
 ($\partial_k A_k^{T}\simeq0$). The potential equation
 is distinguished
 by the Laplacian $\triangle A^T_0$
 without the time derivatives
 in contrast to  the
 dynamic variable equation with
 d'Alambertian $\Box A_k^{T}$. Thus, the second N\"other theorem \cite{Noter}
 means that diffeomorphisms lead to
 the first class constraints removing pure gauge velocities,
 whereas the corresponding pure gauge variables are removed by
 the second class constraints.
 There are similar problems in GR where a reference frame is given as a simplex.

\subsection{\label{adm1}The Dirac -- ADM variables}

  The similar separation of diffeo-invariant dynamic metric components
  from  potentials in GR is fulfilled
  in the specific reference frame
  in terms of the Dirac -- ADM variables \cite{dir}
  defined as the Lichnerowicz transformation to
  the scale-invariant quantities $\omega_{(\mu)}^{(L)}$ \cite{L,Y}
 \be \label{1adm}
 \omega_{(0)}=\psi^6N_{\rm d}dx^0\equiv\psi^2\omega_{(0)}^{(L)},~~~~~~~~~~~
 \omega_{(b)}=\psi^2 {\bf e}_{(b)i}(dx^i+N^i dx^0)
 \equiv\psi^2\omega_{(b)}^{(L)};
 \ee
 here triads ${\bf e}_{(a)i}$ form the spatial metrics with $\det |{\bf
 e}|=1$, $N_{\rm d}$ is the Dirac lapse function, $N^i$ is  shift
 vector and $\psi$ is a determinant of the spatial metric.

 The
 Hilbert action (\ref{gr}) in terms of the  Dirac variables (\ref{1adm})
 takes the form
\be \label{sv111}
 S_{\rm GR}= \int d^4x
 \left[{\bf K}[\vh_0| {e}]-{\bf P}[\vh_0|{e}]+
{\bf S}[\vh_0| {e}]\right],
 \ee
 where
\bea
 {\bf K}[\vh_0|e]&=&{{N}_d}\vh_0^2\left[-{\vphantom{\int}}4
 {  {v}_\psi}^2+\frac{v^2_{(ab)}}{6}\right],
 \label{k11}\\
 {\bf
 P}[\vh_0|e]&=&\frac{{N_d}\varphi_0^2{\psi}^{7}}{6}\left[
 {}^{(3)}R({\bf e}){\psi}+
 {8}\triangle{\psi}\right],
 \label{p11}\\
 {\bf S}[\vh_0|e]&=&2\varphi_0^2\left[\partial_0{{v_\psi}}\right]-
 \partial_j\left[2\varphi^2(N^j {v_\psi})+
 \frac{\varphi_0^2}3
{\psi}^2\partial^j ({\psi}^6
 {N}_d)\right]\label{s11},
 \eea
 are the kinetic, potential, and surface terms,
 respectively,
  \bea\label{proi11}
 {v_\psi}&=&\frac{1}{{N_d}}\left[
 (\partial_0-N^l\partial_l)\log{
 {\psi}}-\frac16\partial_lN^l\right],\\\label{proi12}
 v_{(ab)}&=&\frac{1}{2}\left({\bf e}_{(a)i}v^i_{(b)}+{\bf
 e}_{(b)i}v^i_{(a)}\right),\\\label{proizvod1}
 v_{(a)i}&=&
 \frac{1}{{N_d}}\left[(\partial_0-N^l\partial_l){\bf e}_{(a)i}
+ \frac13 {\bf
 e}_{(a)i}\partial_lN^l-{\bf e}_{(a)l}\partial_iN^l\right]
 \eea
 are velocities of the metric components,
   $\triangle\psi=\partial_i({\bf e}^i_{(a)}{\bf
 e}^j_{(a)}\partial_j\psi)$ is the covariant Laplace operator,
 ${}^{(3)}R({\bf{e}})$ is a three-dimensional curvature
 expressed in terms of triads
   ${\bf e}_{(a)i}$:
\be \label{1-17}
 {}^{(3)}R({\bf e})=-2\partial^{\phantom{f}}_{i}
 [{\bf e}_{(b)}^{i}\sigma_{{(c)|(b)(c)}}]-
 \sigma_{(c)|(b)(c)}\sigma_{(a)|(b)(a)}+
 \sigma_{(c)|(d)(f)}^{\phantom{(f)}}\sigma^{\phantom{(f)}}_{(f)|(d)(c)}.
 \ee
 Here
 \be\label{1-18} \sigma_{(a)|(b)(c)}=
 {\bf e}_{(c)}^{j}
 \nabla_{i}{\bf e}_{(a) k}{\bf e}_{(b)}^{\phantom{r}k}=
 \frac{1}{2}{\bf e}_{(a)j}\left[\partial_{(b)}{\bf e}^j_{(c)}
 -\partial_{(c)}{\bf e}^j_{(b)}\right]
  \ee
  are the coefficients of the spin-connection (see \cite{ll} Eq.
  (98.9)),
  $\nabla_{i}{\bf e}_{(a) j}=\partial_{i}{\bf e}_{(a)j}
  -\Gamma^k_{ij}{\bf e}_{(a) k}$~are covariant derivatives, and
  $\Gamma^k_{ij}=\dfrac{1}{2}{\bf e}^k_{(b)}(\partial_i{\bf e}_{(b)j}
  +\partial_j{\bf e}_{(b)i})$.

 The  definition of canonical momenta
\bea\label{m-1}{p_{\psi}}&=&\frac{\partial {\bf
K}[\vh_0|e]}{\partial
 (\partial_0\ln{{\widetilde{\psi}}})}=-8\vh^2{{v_\psi}},
 \\\label{m-2}
 p^i_{(b)}&=&\frac{\partial {\bf K}[\vh_0|e]
 }{\partial(\partial_0{\bf e}_{(a)i})}
 ={\bf e}^i_{(a)}\frac{\vh^2}{3} v_{(a b)}
 \eea
 allows us to represent
  the action  (\ref{sv111})
   in the Hamiltonian form
 \be\label{h-1}
 S_{\rm GR}=\int dx^0\int d^3x
 \left(\sum\limits_{{F}
 } P_{F}\partial_0F
 +{\cal C}-{N_d} T_{0}^0\right),
 \ee
 where
 $P_F=(p_{\psi},p^i_{(b)})$ are the set of
 momenta (\ref{m-1}) -- (\ref{m-2}),
\be\label{h-3}
 T_{0}^0= {\psi}^{7}\hat \triangle
{\psi}+
  \sum\limits_{I=0,8} {\psi}^I\tau_I,
 \ee
 is the sum of the Hamiltonian densities
 \bea\label{h-4}
 {\psi}^{7}\hat \triangle
 {\psi}&\equiv&{\psi}^{7}\dfrac{4\varphi^2}{3}
 \partial_{(b)}\partial_{(b)}{\psi},\\\label{h-5}
 \tau_{I=0}&=&\dfrac{6{p}_{(ab)}{p}_{(ab)}}{\vh^2}
 -\dfrac{16}{\vh^2}p^2_{\psi},
 \\\label{h-6}
 \tau_{I=8}&=&\dfrac{\varphi^2}
  {6}R^{(3)}({\bf e}),
\eea
 here ${p}_{(ab)}=\dfrac{1}{2}({\bf e}^k_{(a)}
 {p}_{(b)k}+
 {\bf e}^k_{(b)}{p}_{(a)k}),
 $
and
 \be\label{h-2}
 {\cal C}=N_{(b)}
  {T}^0_{(b)} +\lambda_0{p_\psi}+ \lambda_{(a)}\partial_k{\bf e}^k_{(a)}
 \ee
 is a set of the Lagrangian multipliers with
  Dirac constraints, including
 three  first class constraints ${T^0_{(a)}}=0$,
where \bea\label{h-c1}
 {T^0_{(a)}}={T}^0_{i}{\bf e}^i_{(a)}=
  -p_{\psi}\partial_{(a)}
 {\psi}+\frac{1}{6}\partial_{(a)}
 (p_{\psi}{\psi}) +
 2p_{(b)(c)}\sigma_{(b)|(a)(c)}-\partial_{(b)}p_{(b)(a)}
  %+{T^0_{(a)}}_{({\rm m})}~ ,
  \eea
 and fourth second class ones \cite{dir}
 $\partial_k{\bf e}^k_{(a)}=0$, $p_\psi=0$.
 The last constraint means the zero velocity of the spatial volume
   element
\be\label{h-c2}
 p_{{\psi}}=-8\vh_0^2{v_{\psi}}=0 \to
 \partial_0({\psi}^6)
 =\partial_l({\psi}^6 N^l),
 \ee
    and it leads to the positive
   Hamiltonian density   (\ref{h-5}).

   In this case, the equation of motion of the
   spatial determinant takes the potential form
   \be\label{h-c3}
  7N_d{\psi}^{7}\hat \triangle
{\psi}+{\psi}\hat \triangle [N_d{\psi}^{7}]+
  N_d\sum\limits_{I=0,8}I {\psi}^I\tau_I=0.
 \ee
 One can see that in a region of the space, where
 two dynamic variables are absent ${\bf e}_{(a)k}=\delta_{(a)k}$
 (i.e. $\tau_{I=0,8}=0$) there is the Schwarzschild-type
 solution of equations $\delta S/\delta N_d=-T_0^0=0$ and (\ref{h-c3})
 that can be written in the form
 \be\label{h-c4}
 \hat \triangle {\psi}=0,~~~\hat \triangle [N_d{\psi}^{7}]=0
 ~~~\to~~~{\psi}=1+\frac{r_g}{r},~~
 [N_d{\psi}^{7}]=1-\frac{r_g}{r},~~N^k=0
 \ee
 where $r_g$ is the constant of the
 integration given by the boundary conditions.

 One can see that the spatial coordinate diffeomorphisms
 $x_{i} \rightarrow  \tilde x_{i}=\tilde x_{i}(x^0,x_{1},x_{2},x_{3})$
  can be used  in order to fix three graviton momenta by the first
  class constraint (\ref{h-c1}) and remove the corresponding
 conjugate variables (i.e.
  three longitudinal gravitons) by the second class constraint
  $\partial_k{\bf e}^k_{(a)}=0$ in
  complete correspondence with the
   application of the second N\"other theorem  in QED \cite{cj}
    (see Section 2.2).

\subsection{Problems of a diffeo-invariant evolution parameter}

   However, the time coordinate diffeomorphisms
   $x_{0} \rightarrow  \tilde x_{0}=\tilde x_{0}(x^0,x_{1},x_{2},x_{3})$
   violate this QED/GR correspondence, because
   the time first class constraint $T_0^0=0$ fixes not velocity but
   variable
   $\psi$, and its velocity is removed by the second class
   constraint $p_\psi=0$. Moreover,
   the Hamiltonian approach to
   GR is  not invariant with respect to
   the time coordinate transformations. One can see that
   the solution (\ref{h-c4}) violates
 the symmetry of the interval  (\ref{1adm})
 $\omega_{(0)}^{(L)}=\psi^4N_ddx^0$
  %,~$\omega_{(k)}^{(L)}=dx_k$
  with respect to reparameterizations of the
 time-coordinate (\ref{1zel}) $x^0~\to~\widetilde{x}^0=\widetilde{x}^0(x^0)$.

  Zel'manov found \cite{vlad} that a reference frame determined by
   forms (\ref{1adm}) is invariant with respect to
   diffeomorphisms
 \be \label{zel}
 x^0 \rightarrow \tilde x^0=\tilde x^0(x^0);~~~~~
 x_{i} \rightarrow  \tilde x_{i}=\tilde x_{i}(x^0,x_{1},x_{2},x_{3})~,
 \ee
 \be \label{kine}
 \tilde N_d = N_d \frac{dx^0}{d\tilde x^0};~~~~\tilde N^k=N^i
 \frac{\partial \tilde x^k }{\partial x_i}\frac{dx^0}{d\tilde x^0} -
 \frac{\partial \tilde x^k }{\partial x_i}
 \frac{\partial x^i}{\partial \tilde x^0}~.
 \ee
 Only this group of transformations conserves
  a family  of hypersurfaces  $x^0=\rm{const.}$,
  and it is called the {\it``kinemetric''} subgroup of the group of
  general coordinate  transformations.
    The {\it``kinemetric''} subgroup contains only homogeneous
 reparameterizations of the coordinate evolution parameter $(x^0)$
 and three local transformations of the spatial coordinates.
 This means that in
 finite space-time the diffeo-variant quantity $(x^0)$
  and
 the corresponding zero energy (\ref{h-3}) are
 not observable.

 We propose here to solve this problem as in \cite{pp,bpp}, where
  the frame
  (\ref{1adm}) is redefined by pointing out
  diffeo-invariant homogeneous {\it``time-like variable''}
  in accordance with the dimension of the diffeomorphism group
  (\ref{zel}).

\section{Diffeo-invariant formulation of GR}

\subsection{A homogeneous scale factor as evolution parameter}

The Friedmann cosmology and the cosmological perturbation theory
\cite{Lif,kodama} applied as the basic tools
 for analysis of modern observational data in astrophysics and cosmology \cite{bard}
 are the irrefutable arguments in favor of identification of
  such a  diffeo-invariant homogeneous
{\it``evolution parameter''}  with the  cosmological
 scale factor  $a(x_0)$.
 This factor is introduced
 by the scale transformation of the metrics
 $g_{\mu\nu}=a^2(x^0)\widetilde{g}_{\mu\nu}$
and any field $F^{(n)}$ with the conformal weight $(n)$:
 $F^{(n)}=a^n(x_0) \widetilde{F}^{(n)}$.
 In particular, the
   curvature
 \be \label{cur}
 \sqrt{-g}R(g)=a^2\sqrt{-\widetilde{g}}R(\widetilde{g})-6a
 \partial_0\left[{\partial_0a}\sqrt{-\widetilde{g}}~\widetilde{g}^{00}\right]\ee
 can be expressed in terms of
   the new lapse
 function $\widetilde{N}_d$ and spatial determinant $\widetilde{\psi}$ in
 Eq. (\ref{1adm})
 \be \label{lfsd}
 \widetilde{N}_d=[\sqrt{-\overline{g}}~\overline{g}^{00}]^{-1}=a^{2}{N}_d,~~~~~~~~
 \widetilde{\psi}=(\sqrt{a})^{-1}\psi.
 \ee
 In order to keep the number of variables, we identify $\log \sqrt{a}$ with
 the  spatial volume ``averaging'' of $\log{\psi}$
 \be\label{1non1}
 \log \sqrt{a}=\langle \log{\psi}\rangle\equiv\frac{1}{V_0}\int
 d^3x\log{\psi},
 \ee
 where $V_0=\int d^3x  < \infty$ is a finite
 volume. In this case, the new determinant variable $\widetilde{\psi}$
  should be given in the orthogonal class of functions
  satisfying  the identity
 \be\label{non1}
 \int d^3x \log\widetilde{\psi}=\int d^3x \left[\log{\psi}
 -\left\langle{ \log{\psi}}\right\rangle\right]\equiv 0.
 \ee
 One can call these  functions the ``deviations''.
 The operation of  ``deviation''
 $\Pi_{\rm de}\cdot \mu=\overline{\mu}={\mu}-\langle\mu\rangle$
 is orthogonal to
 the operation of  ``averaging''
 $\Pi_{\rm av}\cdot\mu= \langle\mu\rangle$.
 The sum of these two operations is equal to unity
 $\Pi_{\rm de}\cdot \mu+\Pi_{\rm av}\cdot \mu=I\mu$,
 and square of these operations give  them again:
 $\Pi_{\rm de}^2=\Pi_{\rm de}$ and $\Pi_{\rm av}^2=\Pi_{\rm av}$.
 One can see that
 the operations of  ``averaging'' $\langle\mu\rangle$ and
 ``deviation'' $\overline{\mu}={\mu}-\langle\mu\rangle$ have
 properties of
 projection operators.

 Therefore,  the variation of any functional of
 a ``deviation'' $\overline{\mu}={\mu}-\langle\mu\rangle$
  $S[\overline{\mu}]=S[\Pi_{\rm de}\cdot\mu]$  with
 respect to the  ``deviation''
 \be\label{dev}
 \frac{\delta S[\Pi_{\rm de}\cdot\mu]}{\delta(\Pi_{\rm de}\cdot\mu)}=
 \Pi_{\rm de}\cdot\left[\frac{\delta S[\mu]}{\delta\mu}\right]
 \ee
 is a ``deviation''.

  After the scale transformation (\ref{cur}), (\ref{lfsd}) action (\ref{gr})
  takes the form
 \be\label{1gr}
 S[\vh_0]=\widetilde{S}[\vh]-
 \int dx^0 (\partial_0\vh)^2\int \frac{d^3x}{\widetilde{N}_d};
 \ee
 here $\widetilde{S}[\varphi]$
  is the action (\ref{gr})  in
 terms of metrics ${\widetilde{g}}$, where  $\vh_0$ is replaced by
 the running scale $\vh(x^0)=\vh_0a(x^0)$ of all masses  of the
 matter fields.

 The
 energy constraint
 \be\label{nph}
 \frac{\delta S[\vh_0]}{\delta
 \widetilde{N}_d}=-{T}_0^0=
 \frac{(\partial_0\varphi)^2}{\widetilde{N}_d^2}-
 \widetilde{T}_0^0=0
 \ee
  takes the
 algebraic form \cite{pp}, where
 \be
 \widetilde{T}^0_0\equiv-
 \frac{\delta \widetilde{S}[\vh]}{\delta  \widetilde{N}_d}
 \ee
  is the local energy density. This equation has the exact solution
  in both  the local sector
 \be\label{13ec}
  N_{\rm inv}={\langle(\widetilde{N}_d)^{-1} \rangle
 \widetilde{N}_d}=
 \frac{{\left\langle\sqrt{{\widetilde{T}^0_0}}\right\rangle}}
 {\sqrt{{\widetilde{T}^0_0}}},
 \ee
 and the homogeneous one
 \be\label{113ec}
 \left[\frac{d\varphi}{d\zeta}\right]^2\equiv\vh'^2=\rho_{(0)}\equiv
 {\left\langle\sqrt{{\widetilde{T}^0_0}}\right\rangle}^2 ,
 \ee
 where
\be\label{13c}
 \zeta(\varphi_0|\varphi)
 \equiv\int dx^0 \langle{ (\widetilde{N}_d)^{-1}}\rangle^{-1}
 =\pm
 \int_{\vh}^{\vh_0}
 \frac{d\widetilde{\vh}}{{\left\langle
 \sqrt{\widetilde{T}_0^0(\widetilde{\vh})}\right\rangle}}
 \ee
 is a diffeo-invariant time-coordinate and $N_{\rm inv}$ is the diffeo-invariant lapse
 function.

 One can see that in the diffeo-invariant formulation of GR
 there is the almost complete QED/GR correspondence
 of the application of the second N\"other theorem, because
 the energy constraint (\ref{113ec})
 fixes  the homogeneous velocity
 of the cosmic evolution. This fixation
 can be treated as
  the Hubble law in cosmology. The scale factor can be removed
  from the reduced phase space of  diffeo-invariant variables,
  but not from measurable quantities, by the Hamiltonian reduction.

  In Appendix A it was shown that
  the interactions with matter in terms of the scale-invariant
   Lichnerowicz
 variables $F^{(n)}_{(L)}=\psi^{-2n}F^{(n)}$, where
 $n$ is the conformal weight,
   do not contain
 the time derivatives of the spatial determinant.

 \subsection{The diffeo-invariant Hamiltonian  formulation}

 In order to calculate the canonical momenta, let us write the
 Hilbert action (\ref{gr}) in terms of the new Dirac variables (\ref{lfsd})
\be \label{sv11}
 S_{\rm GR}[g=a^2{\widetilde{g}}]= \int d^4x
 \left[{\bf K}[\vh| e]-{\bf P}[\vh|e]+
{\bf S}[\vh| e]\right]-\int
dx^0\frac{(\partial_0\vh)^2}{\widetilde{N}_d}=\int dx^0 L_{\rm
GR},
 \ee
 where
\bea
 {\bf K}[\vh|e]&=&\widetilde{N}_d\vh^2\left[-{\vphantom{\int}}4
 {  \overline{v}}^2+\frac{v^2_{(ab)}}{6}\right],
 \label{k1}\\
 {\bf
 P}[\vh|e]&=&\frac{\widetilde{N}_d\varphi^2\widetilde{\psi}^{7}}{6}\left[
 {}^{(3)}R({\bf e})\widetilde{\psi}+
 {8}\triangle\widetilde{\psi}\right],
 \label{p1}\\
 {\bf S}[\vh|e]&=&2\varphi^2\left[\partial_0{\overline{v}}\right]-
 \partial_j\left[2\varphi^2(N^j \overline{v})+
 \frac{\varphi^2}3
\widetilde{\psi}^2\partial^j (\widetilde{\psi}^6
 {N}_d)\right]\label{s1},
 \eea
 are the kinetic, potential, and ``quasi-surface'' terms,
 respectively,
  \bea\label{proi}
 \overline{v}&=&\frac{1}{\widetilde{N}_d}\left[
 (\partial_0-N^l\partial_l)\log{
 \widetilde{\psi}}-\frac16\partial_lN^l\right],
 \eea
  $v_{(ab)}$ are velocities of the metric components given by Eqs.
 (\ref{proi12}) and (\ref{proizvod1}),
   $\triangle\psi=\partial_i({\bf e}^i_{(a)}{\bf
 e}^j_{(a)}\partial_j\psi)$ is the covariant Laplace operator,
 ${}^{(3)}R({\bf{e}})$ is a three-dimensional curvature
 expressed in terms of triads
   ${\bf e}_{(a)i}$ (\ref{1-17}).

 One can see that the
Lagrangian in the action (\ref{sv11}) includes three terms
describing the spatial metric determinant \be\label{sd}
 L_{\rm GR}=\int d^3x {\cal L}_{\rm GR}=-\int d^3x\overline{{N}_d}
 \left[{\vphantom{\int}}4\vh^2~
 {(\overline{v})}^2
 + 4\vh~ \dfrac{\partial_0\vh}{\widetilde{N}_d}~  \overline{v}+
 \left(\dfrac{\partial_0\vh}{\widetilde{N}_d}\right)^2\right]+...,
 \ee
 where  the first term arises from
 ${\bf K}$  (\ref{k1}), the second one
 (i.e. the velocity  interaction)
 goes from the first
term in ${\bf S}$  (\ref{s1}), and
  the third term goes from the scale factor term in Eq. (\ref{sv11}).

  Keeping
 the number of variables
 $\langle{\log \widetilde{\psi}}\rangle\equiv 0$ (\ref{non1}) and their
 velocities
 \be\label{super2}
 \langle\overline{v}\rangle\equiv 0
 \ee
 means that both $\log \widetilde{\psi}$ and $\overline{v}$ are given   in the class of
 ``deviation'' functions
 distinguished by the projection operator
 $F=\overline{F}-\langle F\rangle$.
   In this class of functions
  the second term in Eq. (\ref{sd}) disappears. In this case momenta (\ref{1pi})
 and  $\overline{p_\psi}$ become
 \be\label{1pi1}
 P_\vh\equiv
  \frac{\partial {L}_{\rm GR}}{\partial(\partial_0 \vh)}=
 -\int d^3x 2\frac{\partial_0\vh}{\widetilde{N}_d}=-2V_0\vh'
 \ee
  and
 \be\label{2pi1}
    \overline{p_{\psi}}\equiv
  \frac{\partial {\cal L}_{\rm GR}}{\partial(\partial_0 \log  \widetilde{\psi})}=
 -8\vh^2 \overline{v}=-\frac{8\vh^2}{\widetilde{N}_d}\left[
 (\partial_0-N^l\partial_l)\log{
 \widetilde{\psi}}-\frac16\partial_lN^l\right].
 \ee
 All velocities   are  expressed in terms of canonical momenta, so that
 the Dirac Hamiltonian approach becomes consistent (see Appendix A).

In the diffeo-invariant version of GR in  finite space-time
 one can
express the action in Hamiltonian form in terms of momenta
(\ref{1pi1}), (\ref{2pi1}) $P_\vh$ and $P_{
F}=[{\overline{p_{\psi}}}, p^i_{{(a)}},p_f]$
 \be\label{hf}
 S[\varphi_0]=\int dx^0\left[\int d^3x \left(\sum_F P_F\partial_0
 F+C-\widetilde{N}_d \widetilde{T}^0_0\right)-P_{\varphi}\partial_0\varphi+
 \frac{P_{\varphi}^2}{4\int d^3x
 (\widetilde{N}_d)^{-1}}\right],
 \ee
 where
 ${\cal C}=N^i {T}^0_{i} +C_0\overline{p_{\psi}}+ C_{(a)}\partial_k{\bf
e}^k_{(a)}$
  is the sum of constraints
  with the Lagrangian multipliers $N^i,C_0,~C_{(a)}$ and the energy--momentum tensor
  components $T^0_i$; these constraints include
   the transversality  $\partial_i {\bf e}^{i}_{(a)}\simeq 0$ and the Dirac
 minimal  surface \cite{dir}
 \be\label{hg}
~~~~~~~~~~~~~~{\overline{p_{\psi}}}\simeq 0 ~~~~\Rightarrow ~~~~
\partial_j(\widetilde{\psi}^6{\cal N}^j)=(\widetilde{\psi}^6)'~~~~~ ({\cal N}^j=N^j\langle
N^{-1}_d\rangle).
 \ee
{\bf The  first class constraints
 including three local ones (\ref{h-c3}) $T_0^i=0$
 and the Hamiltonian version of the
 homogeneous part of the energy constraint} (\ref{113ec})
\be\label{ec}
  P^2_{\varphi}=E^2_{\varphi},
 \ee
 where
 \be\label{hf1}
~~~~~~~~E_\vh=2\int d^3x\sqrt{{\widetilde{T}^0_0}}=
 2V_0{\left\langle\sqrt{{\widetilde{T}^0_0}}\right\rangle},
 \ee
  fix three local longitudinal  momenta and one homogeneous
  momentum $P_\vh$,
 so that the dimension of the first class constraints
 coincides with the dimension of the
kinemetric group of diffeomorphisms (\ref{zel}) of the Hamiltonian
formulation of GR. The similar description of the first class
constraint for the relativistic string is considered in Appendix
B.

\subsection{The Lifshits perturbation theory as an obstacle to
Hamiltonian approach}

Let us consider the definition  of the deviation $\overline{v}$ in
the class of function
 \be\label{friction}
 \langle\overline{v}\rangle \not = 0,
 \ee
that does not keep the number of variable in GR. In this class of
 functions there is the coincidence of the
   homogeneous canonical momentum
 \be\label{1pi}
 P_\vh\equiv
  \frac{\partial {L}_{\rm GR}}{\partial(\partial_0 \vh)}=
 -\int d^3x\left[  4\vh~\overline{v}+2\frac{\partial_0\vh}{\widetilde{N}_d}\right]
 %\equiv -[4\vh V_\psi+2V_\vh]
 \ee
 with the
 zero Fourier harmonics of $\overline{p_\psi}=
 \dfrac{\partial {\cal L}_{\rm GR}}{\partial(\partial_0 \log  \widetilde{\psi})}$
 \be\label{2pi}
   \int d^3x  \overline{p_{\psi}}\equiv \int d^3x
  \frac{\partial {\cal L}_{\rm GR}}{\partial(\partial_0 \log  \widetilde{\psi})}=
 -2\vh \int d^3x \left[4\vh \overline{v}+
  2~\dfrac{\partial_0\vh}{\widetilde{N}_d}~\right]
 \equiv   2\varphi P_{\varphi}.
 \ee
 This means that the velocities could not be expressed in terms of
 canonical momenta and the system with the additional variable
 is not
 the Hamiltonian one. It is just the case of the standard
 cosmological perturbation theory \cite{Lif,kodama,bard}
 based on the interval
 \be\label{L-1}ds^2=
 a^2(\eta)(1+2\Phi)d\eta^2-a^2(\eta)(1-2\Psi)dx^idx^j
 \ee
 for which the Einstein equations
  $$
  2R^\mu_\nu -\delta^\mu_\nu R=4\pi G~ T^\mu_\nu\equiv t^\mu_\nu
  $$
 take the form (see Eq.(4.15) in \cite{bard})
 \bea\label{L-2}
 -3{\cal H}({\cal H}\Phi+\Psi')+\triangle\Psi &= \delta t_{00}\\\label{L-3}
 3[(2{\cal H}'+{\cal H}^2)\Phi+{\cal H}\Phi'+\Psi''+2{\cal
 H}\Psi']+\triangle(\Phi-\Psi)&=\delta t_{ii},
 \eea
 where ${\cal H}=a'/a$, and $\delta t_{\mu \nu}$ are the perturbations.
  These equations follow from the variational principle, if the
  correspondent action of the type of (\ref{sv11})
  (considered in \cite{bard}, see Eq. (10.7) p. 261)
   contains the  velocity-velocity interaction.
 One can be convinced that in this case
the canonical momenta $P_a~~\equiv~~
  \dfrac{\partial {L}_{GR}}{\partial(\partial_0 a)}$ and $P_\Psi\equiv
 \left\langle\dfrac{\partial{\cal L}_{GR}}{\partial(\partial_0 {\Psi})}\right\rangle$
coincide $2aP_a=P_\Psi$.
  The Hamiltonian approach is failure, as
``velocities''
  $V_a=\partial_0 a$ and
  $V_\Psi=\left\langle\partial_0\Psi\right\rangle$ could not
  be expressed in terms of $P_a,P_\Psi$.
 The strong constraints
 $
 \left\langle\Psi \right\rangle\equiv0;~\left\langle\partial_0\Psi \right\rangle\equiv0
 $
 return us back to the Hamiltonian GR.

As we  see below, the main differences of the
 ``friction-free'' version of GR from the Lifshits version \cite{Lif,bard} are the
 potential perturbations of the scalar components ${N_{\rm inv}},
 \widetilde{\psi}$ (given by Eqs. (\ref{2-17}) --
(\ref{2-23}) in Section 4)
 instead of the kinetic ones and the
 nonzero shift vector $N^k\not =0$ (determined by Eq. (\ref{La-6})).
  Recall that just the kinetic perturbations
   are responsible for the ``primordial power spectrum'' in
  the inflationary model \cite{bard}. The problem appears to
  describe CMBR by the potential perturbations.

\subsection{The  Hamiltonian reduction}

One can find the
 evolution of all field variables $F(\vh,x^i)$  with respect to
 $\vh$ by variation of the ``reduced'' action obtained as
   values of the Hamiltonian form of the initial action  (\ref{hf})
 onto the energy constraint (\ref{ec})
 \be\label{2ha2} S[\vh_0]{}_{{}_{{P_\vh=\pm E_\vh}}} =
 \int\limits_{\vh}^{\vh_0}d\widetilde{\vh} \left\{\int d^3x
 \left[\sum\limits_{  F}P_{  F}\partial_\vh F
 +\bar{\cal C}\mp2\sqrt{\widetilde{T}_0^0(\widetilde{\vh})}\right]\right\},
\ee
 where $\bar{\cal C}={\cal
 C}/\partial_0\widetilde{\vh}$ \cite{pp}.

 Here the reduced Hamiltonian function given by Eq. (\ref{hf1})
 can be treated as the ``universe energy'' by analogy with the ``particle energy'' in
 special relativity (SR).
   The reduced Hamiltonian $\sqrt{\widetilde{T}_0^0}$ is
 Hermitian,
 as the  minimal surface
 constraint
 (\ref{hg}) removes a negative
 contribution of $\overline{p_{\psi}}$ from energy density.
Thus, the diffeo-invariance   gives us the
 solution of the problem of nonzero energy  in GR by the
 Hamiltonian reduction like  solution of a similar problem in
 SR.

One can see, that in the diffeo-invariant formulation of GR
considered here the Hubble parameter becomes the generator of
evolution with respect to the dynamic evolution parameter that
abandons the reduced phase space but not the set of observables. The
similar solution of the problem of the energy for the relativistic
string is considered in Appendix B.

 The main consequence of the separation of the cosmological  scale
 factor is the  globalization of the
  energy constraint (\ref{ec}). It fixes only
  the scale momentum ${P_\vh}_{\pm}=\pm E_\vh,$ the values of which
   become  the generator of evolution of all variables with respect to the
  evolution parameter $\vh$ \cite{pp} forward and backward,
 respectively. The negative energy problem can be solved
 by analogy with
 the modern quantum field theory  as
 the primary quantization of the energy constraint
 $[{P^2_\vh}-E^2_\vh]\Psi_{\rm u}=0$ and the secondary quantization
 $\Psi_{\rm u}=(1/\sqrt{2E_\vh)}[A^++A^-]$ by the Bogoliubov
 transformation $ A^+=\alpha
 B^+\!+\!\beta^*B^-$, in order to diagonalize the equations of
 motion by the condensation of ``universes''
 $<0|\dfrac{i}{2}[A^+A^+-A^-A^-]|0>=R(\vh)$
 and describe  cosmological creation of a  ``number'' of universes
  $<0|A^+A^-|0>=N(\vh)$
  from the stable Bogoliubov vacuum  $B^-|0>=0$ \cite{origin}.
 Vacuum postulate $B^-|0>=0$ leads to an arrow of the invariant
time $\zeta\geq 0$ (\ref{13c}) and its absolute point of reference
$\zeta= 0$ at the moment of creation $\vh=\vh_I$; whereas the
Planck value of
 the running mass scale $\vh_0=\vh(\zeta=\zeta_0)$ belongs to the present
 day moment $\zeta_0$.

 The reduced action (\ref{2ha2}) shows us
 that the initial data at the beginning $\vh=\vh_I$ are independent of
  the present-day ones at  $\vh=\vh_0$,
  therefore
  the proposal about an existence of the  Planck epoch $\vh=\vh_0$
   at the beginning \cite{bard} looks
  very doubtful.

 \section{The Diffeo-Invariant  Scalar Potential Perturbations}

\subsection{The Hamiltonian perturbation theory}

 In diffeo-invariant formulation of GR in the specific
  reference frame the scalar potential perturbations
  can be defined as $N^{-1}_{\rm int}=1+\overline{\nu}$ and
 $\widetilde{\psi}=e^{\overline{\mu}}=1+\overline{\mu}+...$, where
 $\overline{\mu},\overline{\nu}$ are given in the class of
 functions distinguished by the projection operator
 $\overline{F}=F-\langle F\rangle$,
 so that $\langle \overline{F}\rangle \equiv 0$.

 The explicit dependence of the metric simplex and the energy
 tensor
 $\widetilde{T}^0_0$ on $\widetilde{\psi}$
  can be given in terms of the scale-invariant Lichnerowicz variables \cite{L}
  introduced in Appendix A (\ref{1-21}) and
 \bea\label{adm-2}
 \omega^{(L)}_{(0)}&=&\widetilde{\psi}^4N_{\rm int}d\zeta,~~~~~
 \omega^{(L)}_{(b)}={\bf e}_{(b)k}[dx^k +{\cal N}^kd\zeta],
 \\
\label{La-2}
 \widetilde{T}^0_0&=& \widetilde{\psi}^{7}\hat \triangle
 \widetilde{\psi}+
  \sum\limits_{I} \widetilde{\psi}^I a^{\frac{I}{2}-2}\tau_I,
  ~~~~~~\tau_I\equiv\langle\tau_I\rangle+\overline{\tau_I},
 \eea
where $\hat \triangle
 \widetilde{\psi}\equiv\dfrac{4\varphi^2}{3}\partial_{(b)}
 \partial_{(b)}\widetilde{\psi}$ is the
 Laplace operator and  $\tau_I$ is partial energy density
  marked by the index $I$ running a set of values
   $I=0$ (stiff), 4 (radiation), 6 (mass), and 8 (curvature)
 in correspondence with a type of matter field contributions
 considered in Appendix A (\ref{h32}) -- (\ref{h35})
 (except of the $\Lambda$-term, $I=12$).
 The negative contribution $-({16}/{\vh^2})\overline{p_{\psi}}^2$ of the
 spatial determinant momentum  in the energy
 density $\tau_{I=0}$
can be removed by the Dirac constraint \cite{dir} of the
 zero velocity of the spatial volume element (\ref{hg})
 \be\label{La-6}
 \overline{p_{\psi}}=
 -8\vh^2\frac{\partial_{\zeta}\widetilde{\psi}^6-\partial_l
 [\widetilde{\psi}^6{\cal N}^l]}{\widetilde{\psi}^6{N}_{\rm int}}=0.
 \ee
 The diffeo-invariant part of the lapse function ${N}_{\rm int}$
 is determined by the local part (\ref{13ec})
 of the energy constraint (\ref{nph}) that can be
 written as
 \be\label{ec1}
 \widetilde{T}^0_0={N}^{-2}_{\rm int}\rho_{(0)}, ~~~~\to
 ~~~~{N}^{-1}_{\rm int}=\sqrt{\widetilde{T}^0_0}\rho_{(0)}^{-1/2},
 \ee
where $\rho_{(0)}=\left\langle
\sqrt{\widetilde{T}^0_0}\right\rangle^2$\!\!.
 In the class of functions $\overline{F}=F-\langle F\rangle$, the classical
 equation ${\delta S}/{\delta \log \widetilde{\psi}}=0$ takes the
 form
 \be\nonumber
 \widetilde{N_d}\widetilde{\psi}
 \frac{\partial \widetilde{T}_0^0}{\partial \widetilde{\psi}}+
 \widetilde{\psi}\triangle \left[
 \frac{\partial \widetilde{T}_0^0}{\partial
 \triangle\widetilde{\psi}}\widetilde{N_d}\right]=0.
 \ee
 Using the property of the deviation projection operator (\ref{dev})
 $\delta S/\delta\overline{\mu}=\overline{D}=D-\langle D\rangle$,
 where $\overline{\mu}=\log \widetilde{\psi}$, we got
 the following equation
\be\label{e2}
 7{N_{\rm inv}\widetilde{\psi}^7\hat \triangle\widetilde{\psi}}
 \!+\!{\widetilde{\psi}\hat \triangle[N_{\rm inv}\widetilde{\psi}^7]}
 \!+\!{N_{\rm inv}}\sum\limits_{I} I{\widetilde{\psi}^I
 a^{\frac{I}{2}-2}\tau_I}=\rho_{(1)},
 \ee
 where $\rho_{(1)}=\left\langle 7{N_{\rm inv}\widetilde{\psi}^7\hat \triangle\widetilde{\psi}}
 \!+\!{\widetilde{\psi}\hat \triangle[N_{\rm inv}\widetilde{\psi}^7]}
 \!+\!\sum\limits_{I} I{\widetilde{\psi}^I
 a^{\frac{I}{2}-2}\tau_I}\right\rangle$. Using (\ref{ec1})
 we can write for $\widetilde{\psi}$ a nonlinear equation
 \be\label{nl}
 ({{\widetilde{T}^0_0})^{-1/2}
 \left[7\widetilde{\psi}^7\hat \triangle\widetilde{\psi}
 \!+\sum\limits_{I} I\widetilde{\psi}^I
 a^{\frac{I}{2}-2}\tau_I \right]+
 \widetilde{\psi}\hat \triangle[({\widetilde{T}^0_0})^{-1/2}\widetilde{\psi}^7]}
 =\rho_{(1)}\rho_{(0)}^{-1/2}.
 \ee

 In the infinite volume limit $\rho_{(n)}=0,~a=1$
 Eqs.  (\ref{ec1}) and (\ref{e2})  coincide with the equations
 of the diffeo-variant formulation of GR
 $T^0_0=0$ and (\ref{h-c3}) considered in Section 2.3.

 For the small deviations $N^{-1}_{\rm int}=1+\overline{\nu}$ and
 $\widetilde{\psi}=e^{\overline{\mu}}=1+\overline{\mu}+...$ the
 first orders of Eqs.  (\ref{ec1}) and (\ref{e2}) take the form
 \bea\label{1e1-2}
   (-\hat \triangle-\rho_{(1)})\overline{\mu}~~~~ +&
   2\rho_{(0)}\overline{\nu}&=~~~\overline{\tau}_{(0)},
 \\\label{1ec1-2}
 (14\hat \triangle+\rho_{(2)})\overline{\mu}~~
 -&~~~~~(\hat
 \triangle+\rho_{(1)})\overline{\nu}&=-~\overline{\tau}_{(1)},
 \eea
 where
 \bea\label{ec1-3}
 \rho_{(n)}=\langle\tau_{(n)}\rangle\equiv\sum_II^na^{\frac{I}{2}-2}\langle\tau_{I}\rangle\\
\label{ec1-4} \tau_{(n)}=\sum_II^na^{\frac{I}{2}-2}\tau_{I}.
 \eea

 The set of Eqs. (\ref{e1-2}) and (\ref{ec1-2})
 gives $\overline{\nu}$ and $\overline{\mu}$ in the form of a sum
  \bea\label{2-17}
 {\overline{\mu}}&=&\frac{1}{14\beta}\int d^3y\left[D_{(+)}(x,y) \overline{T_{(+)}}(y)-
 D_{(-)}(x,y) \overline{T_{(-)}}(y)\right],\\\label{2-18}
 {\overline{\nu}}&=&\frac{1}{2\beta}\int d^3y\left[(1+\beta)D_{(+)}(x,y) \overline{T_{(+)}}(y)-
 (1-\beta)D_{(-)}(x,y) \overline{T_{(-)}}(y)\right],
  \eea
 where
 \be\label{beta}
 \beta=\sqrt{1+[\langle \tau_{(2)}\rangle-14\langle
\tau_{(1)}\rangle]/(98\langle \tau_{(0)}\rangle)}, \ee
 \be\label{cur1}
 \overline{T_{(\pm)}}=(7\overline{\tau_{(0)}}-
 \overline{\tau_{(1)}})~\pm~ 7\beta\overline{\tau_{(0)}}
 %=7(1\pm\beta)\overline{\tau_{(0)}}-\overline{\tau_{(1)}}
 \ee
 are the local currents, $D_{(\pm)}(x,y)$ are the Green functions satisfying
 the equations
 \bea\label{2-19}
 [\pm \hat m^2_{(\pm)}-\hat \triangle
 ]D_{(\pm)}(x,y)=\delta^3(x-y),
 \eea
 where $\hat m^2_{(\pm)}= 14 (\beta\pm 1)\langle \tau_{(0)}\rangle \mp
\langle \tau_{(1)}\rangle$.

The reduced Hamiltonian
 function (\ref{hf1}) in terms of this solutions takes the form of the current-current interaction
\bea\label{hf2} E_\vh&=&2\int d^3x\sqrt{{T^0_0}}=
2V_0\sqrt{\langle \tau_{(0)}\rangle}+\\\nonumber
 &+&\frac{1}{14\beta
  \sqrt{\langle\tau_{(0)}\rangle}}
 \int d^3x \int d^3y \left[\overline{T_{(+)}}(x)D_{(+)}(x,y)
 \overline{T_{(+)}}(y)
 +\overline{T_{(-)}}(x)D_{(-)}(x,y) \overline{T_{(-)}}(y)\right].
 \eea
   In the case of point mass distribution in a finite volume $V_0$ with the zero pressure
  and  the  density
  $\overline{\tau_{(1)}}=\dfrac{\overline{\tau_{(2)}}}{6}
  \equiv \sum\limits_{J} M_J\left[\delta^3(x-y_J)-\dfrac{1}{V_0}\right]$,
 solutions   (\ref{2-17}),  (\ref{2-18}) take
 a very important form
 \bea\label{2-21}
  \overline{\mu}(x)&=\sum\limits_{J}
  \dfrac{r_{gJ}}{4r_{J}}\left[{\gamma_1}e^{-m_{(+)}(z)
 r_{J}}+ (1-\gamma_1)\cos{m_{(-)}(z)
 r_{J}}\right],\\\label{2-22}
 \overline{\nu}(x)&=\sum\limits_{J}
 \dfrac{2r_{gJ}}{r_{J}}\left[(1-\gamma_2)e^{-m_{(+)}(z)
 r_{J}}+ {\gamma_2}\cos{m_{(-)}(z)
 r_{J}}\right],
 \eea
 where
 $$
  {\gamma_1}=\frac{1+7\beta}{14\beta},~~~
 {\gamma_2}=\frac{(1-\beta)(7\beta-1)}{16\beta},~~
 r_{gJ}=\frac{3M_J}{4\pi\vh^2},~~
 r_{J}=|x-y_J|,~~~~m^2_{(\pm)}=\hat m^2_{(\pm)}\frac{3}{4\vh^2}.
 $$
 The minimal surface (\ref{hg})
  $\partial_i[\overline{\psi}^6{\cal N}^i]-(\overline{\psi}^6)'=0$
 gives the shift of the coordinate
  origin in the process of evolution
 \be \label{2-23}
{\cal
 N}^i=\left(\frac{x^i}{r}\right)\left(\frac{\partial_\zeta V}{\partial_r V}\right),~~~
 ~~~V(\zeta,r)=\int\limits_{0}^{r}d\widetilde{r}
 ~\widetilde{r}^2\widetilde{\psi}^6(\zeta,\widetilde{r}).
  \ee
In the infinite volume limit $\langle \tau_{(n)}\rangle=0$ these
 solutions take the standard Newtonian form:
 $\overline{\mu}=D\cdot \tau_{(0)}$, $\overline{\nu}=D\cdot [14\tau_{(0)}-\tau_{(1)}]$,
 ${\cal N}^i=0$
 (where $\hat \triangle D(x)=-\delta^3(x)$).

 \subsection{Perturbation theory as generalization of  Schwarzschild solution}
One can see that another choice of variables for
 scalar potentials rearranges the perturbation series and leads
 to another result. In order to demonstrate this fact, let us
 choose the lapse function (\ref{13ec}) as
 ${N_{\rm inv}}\widetilde{\psi}^{7}
 =1-\overline{\nu_1}$ and keep
 $\widetilde{\psi}=1+\overline{\mu_1}$.
In order to simplify equations of the scalar potentials
  ${N}_{\rm int},\widetilde{\psi}$, one can introduce new
 table of symbols:
 \be\label{nts}
 N_{\rm s}=\psi^7 N_{\rm inv}, ~~~~T(\widetilde{\psi})=
 \sum\limits_{I} \widetilde{\psi}^{(I-7)}
 a^{\frac{I}{2}-2}\tau_I,~~~~~~\rho_{(0)}=\left\langle
\sqrt{\widetilde{T}^0_0}\right\rangle^2=\vh'^2.
 \ee
 In terms of these symbols the action (\ref{hf}) can be presented
 as a generating functional of equations of the local scalar potentials
 $N_{\rm s},\widetilde{\psi}$ and  field variables $F$ in terms of
 diffeo-invariant time $\zeta$:
 \be\label{gf}
 S[\varphi_0]=\int d\zeta\int d^3x \left[\sum_F P_F\partial_\zeta
 F+C_\zeta-{N}_{\rm s}\left(\hat \triangle
 \widetilde{\psi}+{T}(\widetilde{\psi})\right)-\frac{\widetilde{\psi}^7
 \rho_{(0)}}{N_{\rm s}}\right],
 \ee
 where $C_\zeta=C (dx^0/d\zeta)$.

%\subsection{The potential equations}

 The variations of this action with respect to
 $N_{\rm s},\widetilde{\psi}$ lead to equations
 \bea\label{4-1}
\hat \triangle
 \widetilde{\psi}+{T}(\widetilde{\psi})&=&\frac{\widetilde{\psi}^7
 \rho_{(0)}}{N^2_{\rm s}},
 \\\label{4-2}
\widetilde{\psi}\hat \triangle{N}_{\rm s}
 +{N}_{\rm s}\widetilde{\psi}\partial_{\widetilde{\psi}}{T}
 +7\frac{\widetilde{\psi}^7
 \rho_{(0)}}{N_{\rm s}}&=&\rho_{(1)},
 \eea
 respectively, we have used here the constraint  (\ref{La-6})
 and the deviation projection operator (\ref{dev}) according
 to which $\rho_{(1)}=\langle\widetilde{\psi}\hat \triangle{N}_{\rm s}
 +{N}_{\rm s}\widetilde{\psi}\partial_{\widetilde{\psi}}{T}
 +7{\widetilde{\psi}^7
 \rho_{(0)}}/{N_{\rm s}}\rangle$.

One can see that in the infinite volume limit
$\rho_{(n)}=\langle\tau_I\rangle=0$ Eqs. (\ref{4-1}) and
(\ref{4-2}) reduce to the  equations of the conventional GR with
the  Schwarzschild solutions
  ${\overline{\psi}}=1+\frac{r_g}{4r};~~
  N_{\rm s}
  =1-\frac{r_g}{4r}$ in empty space, where Eqs. (\ref{4-1}) and
(\ref{4-2}) become
  $\hat \triangle {\overline{\psi}}=0, ~\hat \triangle N_{\rm
  s}=0$.

%\subsection{Cosmological perturbations}

 For the small deviations $N_{\rm s}=1-{\nu}_1$ and
 $\widetilde{\psi}=1+{\mu}_1$ the
 first orders of Eqs.  (\ref{4-1}) and
(\ref{4-2}) take the form
 \bea\label{e1-2}
   [-\hat{\triangle}+14\rho_{(0)}-\rho_{(1)}]\mu_{1}~~ +&
   2\rho_{(0)}\nu_1&=~~~\overline{\tau}_{(0)}
 \\\label{ec1-2}
 [7\cdot 14\rho_{(0)}-14\rho_{(1)}+\rho_{(2)}]\mu_1~~
 +&[-\hat{\triangle}+
14\rho_{(0)}-\rho_{(1)}]\nu_1&=7\overline{\tau}_{(0)}-\overline{\tau}_{(1)},
 \eea
%$$[-\hat{\triangle}+14\rho_{(0)}-\rho_{(1)}]\mu_{1}+2\rho_{(0)}\nu_1=
%\tau_{(0)}$$
%
%$$[7\cdot 14\rho_{(0)}-14\rho_{(1)}+\rho_{(2)}]\mu_1+[(-\hat{\triangle}+
%14\rho_{(0)})-\rho_{(1)}]\nu_1=(7\bar{\tau}_{(0)}-\bar{\tau}_{(1)})$$
where
 \bea\label{1ec1-3}
 \rho_{(n)}=\langle\tau_{(n)}\rangle
 \equiv\sum_II^na^{\frac{I}{2}-2}\langle\tau_{I}\rangle.%\\
%\label{ec1-4} \tau_{(n)}=\sum_II^na^{\frac{I}{2}-2}\tau_{I}.
 \eea

%%%%%%%%%%%%%%%%%%%%%%%%%%%%%%%%%%

 This choice of variables
 determines $\overline{\mu_1}$ and $\overline{\nu_1}$ in the form of a sum
 %\cite{origin}
  \bea\label{12-17}
 \widetilde{\psi}=1+{\overline{\mu_1}}&=&1+\frac{1}{2}\int d^3y\left[D_{(+)}(x,y)
\overline{T}_{(+)}^{(\mu)}(y)+
 D_{(-)}(x,y) \overline{T}^{(\mu)}_{(-)}(y)\right],\\\label{12-18}
 N_{\rm inv}\widetilde{\psi}^7=1-{\overline{\nu_1}}&=&1-\frac{1}{2}\int d^3y\left[D_{(+)}(x,y)
\overline{T}^{(\nu)}_{(+)}(y)+
 D_{(-)}(x,y) \overline{T}^{(\nu)}_{(-)}(y)\right],
  \eea
 where $\beta$ are given by Eqs. (\ref{beta})
 \be\label{1cur1}\overline{T}^{(\mu)}_{(\pm)}=\overline{\tau_{(0)}}\mp7\beta
  [7\overline{\tau_{(0)}}-\overline{\tau_{(1)}}],
 ~~~~~~~
 \overline{T}^{(\nu)}_{(\pm)}=[7\overline{\tau_{(0)}}-\overline{\tau_{(1)}}]
 \pm(14\beta)^{-1}\overline{\tau_{(0)}}
 \ee
 are the local currents, $D_{(\pm)}(x,y)$ are the Green functions satisfying
 the equations (\ref{2-19})
 where $\hat m^2_{(\pm)}= 14 (\beta\pm 1)\langle \tau_{(0)}\rangle \mp
\langle \tau_{(1)}\rangle$. In the finite volume limit these
solutions for $\widetilde{\psi}, N_{\rm inv}$ coincide with
solutions (\ref{2-17}) and (\ref{2-18}), where
$\overline{\nu_1}=\overline{\nu} -7\overline{\mu}$ and
$\overline{\mu_1}=\overline{\mu}$.

  In the case of point mass distribution in a finite volume $V_0$ with the
zero pressure
  and  the  density
  $\overline{\tau_{(0)}}(x)=\dfrac{\overline{\tau_{(1)}}(x)}{6}
  \equiv  M\left[\delta^3(x-y)-\dfrac{1}{V_0}\right]$,
 solutions   (\ref{12-17}),  (\ref{12-18}) take
 a  form
 \bea\label{12-21}
  \widetilde{\psi}&=1+
  \dfrac{r_{g}}{4r}\left[{\gamma_1}e^{-m_{(+)}(z)
 r}+ (1-\gamma_1)\cos{m_{(-)}(z)
 r}\right],\\\label{12-22}
 N_{\rm inv}\widetilde{\psi}^{7}&=1-
 \dfrac{r_{g}}{4r}\left[(1-\gamma_2)e^{-m_{(+)}(z)
 r}+ {\gamma_2}\cos{m_{(-)}(z)
 r}\right],
 \eea
 where
 $
  {\gamma_1}=\dfrac{1+7\beta}{2},~~~
 {\gamma_2}=\dfrac{14\beta-1}{28\beta},~~
 r_{g}=\dfrac{3M}{4\pi\vh^2},~~
 r=|x-y|.
 $
 Both choices of variables  (\ref{2-21}),  (\ref{2-22}) and
(\ref{12-21}),  (\ref{12-22}) have spatial oscillations and the
nonzero shift of the coordinate
  origin of the type of (\ref{2-23}).

In the infinite volume limit $\langle \tau_{(n)}\rangle=0,~a=1$
 solutions (\ref{12-21}) and  (\ref{12-22}) coincide with
 the isotropic version of  the Schwarzschild solutions:
 $\widetilde{\psi}=1+\dfrac{r_g}{4r}$,~
 ${N_{\rm inv}}\widetilde{\psi}^{7}=1-\dfrac{r_g}{4r}$,~$N^k=0$.
 It is of interest  to find  an exact solution of Eq. (\ref{nl}) for
 different equations of state.

\section{Cosmic Microwave Background Radiation}

\subsection{Status of  SN data in terms of the scale-invariant variables}

 Einstein's correspondence principle \cite{pp}
 as the low-energy
 expansion of the {\it``reduced action''} (\ref{2ha2}) over the
 field density ${T}_{{\rm s}}$
 $$2d\vh \sqrt{T_0^0}= 2d\vh
 \sqrt{\rho_{0}(\vh)+{T}_{{\rm s}}}
 =
 d\vh
 \left[2\sqrt{\rho_0(\vh)}+
 {{T}_{{\rm s}}}/{\sqrt{\rho_0(\vh)}}\right]+...$$
 gives the sum:
 $S^{(+)}|_{\rm
 constraint}= S^{(+)}_{\rm cosmic}+S^{(+)}_{\rm
 field}+\ldots$, where
 $S^{(+)}_{\rm cosmic}[\varphi_I|\varphi_0]= -
 2V_0\int\limits_{\vh_I}^{\vh_0}\!
 d\vh\!\sqrt{\rho_0(\vh)}$ is the reduced  cosmological action (\ref{2ha2}),
 and
 \be\label{12h5} S^{(+)}_{\rm field}=
 \int\limits_{\eta_I}^{\eta_0} d\eta\int\limits_{V_0}^{} d^3x
 \left[\sum\limits_{ F}P_{ F}\partial_\eta F
 +\bar{{\cal C}}-{T}_{{\rm s}}\right]
 \ee
 is the standard field action
 in terms of the conformal time:
 $d\eta=d\vh/\sqrt{\rho_0(\vh)}$,
 in the conformal flat space--time with running masses
 $m(\eta)=a(\eta)m_0$.

 \begin{figure}[t]
\vspace{1cm}
 \begin{center}
 \includegraphics[width=0.65\textwidth,clip]{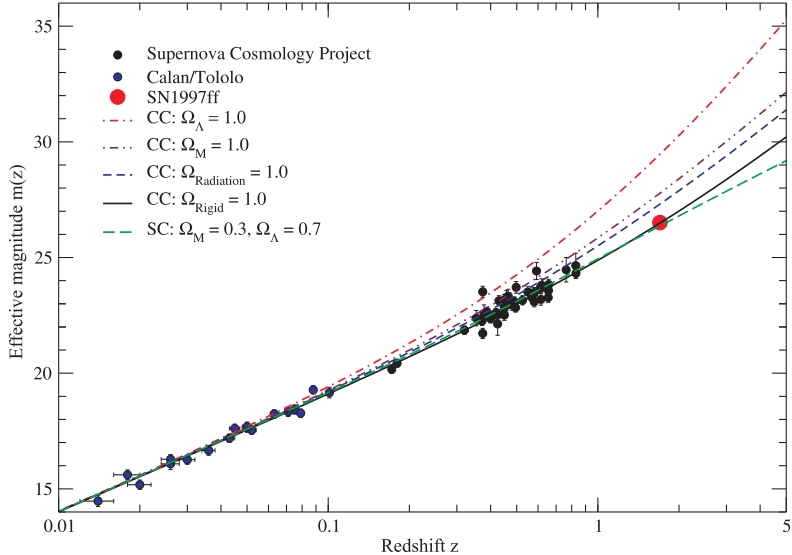}
\caption{ The Hubble diagram~\cite{039a,Danilo} in cases of the
{\it``scale-variant''} units of standard cosmology (SC)  and the
{\it``scale-invariant''} ones of conformal cosmology (CC).
 The points include  42 high-redshift Type Ia
 supernovae~\protect\cite{snov} and the reported
 farthest supernova SN1997ff~\protect\cite{SN}. The best
fit to these data  requires a cosmological constant
$\Omega_{\Lambda}=0.7$, $\Omega_{\rm CDM}=0.3$ in the case of SC,
whereas in CC
 these data are consistent with  the dominance of the rigid (stiff) state.
\label{fig1}}
\end{center}
\end{figure}

 This expansion shows us that the Hamiltonian approach
 in terms of the Lichnerowicz scale-invariant variables
 (\ref{adm-2}) and (\ref{1-21})
 identifies the ``conformal quantities''
  with the observable ones including the conformal time $d\eta$,
  instead of $dt=a(\eta)d\eta$, the coordinate
 distance $r$, instead of Friedmann one $R=a(\eta)r$, and the conformal
 temperature $T_c=Ta(\eta)$, instead of the standard one $T$.
 Therefore,
 the scale-invariant variables  distinguish the conformal cosmology (CC)
 \cite{039,Narlikar},
  instead of the standard cosmology (SC).
 In this case,
 the
  red shift of the spectral lines of atoms on cosmic objects
 $$
\frac{E_{\rm emission}}{E_0}=\frac{m_{\rm atom}(\eta_0-r)}{m_{\rm
atom}(\eta_0)}\equiv\frac{\vh(\eta_0-r)}{\vh_0}=a(\eta_0-r)
=\frac{1}{1+z}
$$
is explained by the running masses $m=a(\eta)m_0$ in action
(\ref{12h5}).

The conformal observable distance  $r$ loses the factor $a$, in
comparison with the nonconformal one $R=ar$. Therefore, in the
 case of CC, the redshift --
  coordinate-distance relation $d\eta=d\vh/\sqrt{\rho_0(\vh)}$
  corresponds to a different
  equation
  of state than in the case of SC \cite{039}.
 The best fit to the data,  including
  Type Ia supernovae~\protect\cite{snov,SN},
 requires a cosmological constant $\Omega_{\Lambda}=0.7$,
$\Omega_{\rm CDM}=0.3$ in the case of the ``scale-variant
quantities`` of standard cosmology. In the case of ``conformal
 quantities'' in CC, the Supernova data \cite{snov,SN} are
consistent with the dominance of the stiff (rigid) state,
$\Omega_{\rm Rigid}\simeq 0.85 \pm 0.15$, $\Omega_{\rm
Matter}=0.15 \pm 0.15$ \cite{039,039a,Danilo}. If $\Omega_{\rm
Rigid}=1$, we have the square root dependence of the scale factor
on conformal time $a(\eta)=\sqrt{1+2H_0(\eta-\eta_0)}$. Just this
time dependence of the scale factor on
 the measurable time (here -- conformal one) is used for description of
 the primordial nucleosynthesis \cite{Danilo,three}.

%%%%%%%%%%%%%%%%%%%%

 This stiff state is formed by a free scalar field
 when $E_\vh=2V_0\sqrt{\rho_0}=Q/\vh$. In this case there is an exact
solution of  Bogoliubov's equations  of the number of universes
created from a vacuum with the initial data
$\vh(\eta=0)=\vh_I,H(\eta=0)=H_I$ \cite{origin}.

\subsection{Cosmological creation of matter}

 These initial data $\vh_I$ and $H_I$ are determined by the
 parameters of matter cosmologically created from the Bogoliubov
 vacuum  at the beginning of a universe $\eta\simeq 0$.

\begin{figure}
  \centering
  \includegraphics{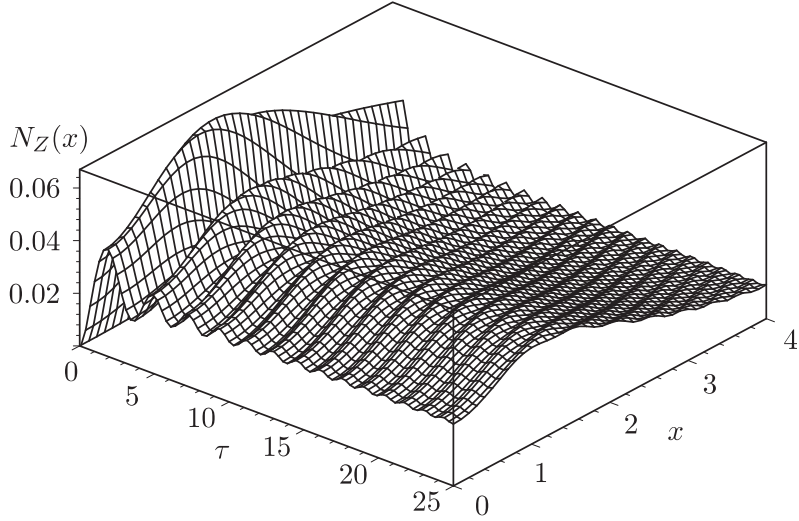}
  \caption{Longitudinal ($N_Z(x)$) components
of the boson distribution versus the dimensionless time $\tau=
2\eta H_I$ and the dimensionless momentum $x = q/M_I$ at the
initial data $M_I = H_I$  \cite{114:a,vin}.\label{fig3}}
\end{figure}

 The Standard
 Model (SM) density ${T}_{{\rm s}}$ in action (\ref{12h5})
  shows
 us that W-, Z- vector bosons have maximal probability of this
 cosmological creation
 due to their mass singularity~\cite{114:a}. One can introduce the notion of
 a particle in a universe if the Compton length of a particle
 defined by its inverse mass
 $M^{-1}_{\rm I}=(a_{\rm I} M_{\rm W})^{-1}$ is less than the
 universe horizon defined by the inverse Hubble parameter
 $H_{\rm I}^{-1}=a^2_{\rm I} (H_{0})^{-1}$ in the
 stiff state. Equating these quantities $M_{\rm I}=H_{\rm I}$
 one can estimate the initial data of the scale factor
 $a_{\rm I}^2=(H_0/M_{\rm W})^{2/3}=10^{-29}$ and the primordial Hubble parameter
 $H_{\rm I}=10^{29}H_0\sim 1~{\rm mm}^{-1}\sim 3 K$.
 Just at this moment there is  an effect of intensive
  cosmological creation of the vector bosons described in
  \cite{114:a,vin} (see Fig. \ref{fig3});
 in particular, the distribution functions of the longitudinal   vector bosons
demonstrate us a large contribution of relativistic momenta.
 Their conformal (i.e. observable) temperature $T_c$
 (appearing as  a consequence of
 collision and scattering of these bosons) can be estimated
from the equation in the kinetic theory for the time of
establishment of this temperature $ \eta^{-1}_{relaxation}\sim
n(T_c)\times \sigma \sim H $, where $n(T_c)\sim T_c^3$ and $\sigma
\sim 1/M^2$ is the cross-section. This kinetic equation and values
of the initial data $M_{\rm I} = H_{\rm I}$ give the temperature
of relativistic bosons
\be\label{t}
 T_c\sim (M_{\rm I}^2H_{\rm I})^{1/3}=(M_0^2H_0)^{1/3}\sim 3 K
\ee
as a conserved number of cosmic evolution compatible with the
Supernova data \cite{039,snov,SN}.
 We can see that
this  value is surprisingly close to the observed temperature of
the CMB radiation
 $ T_c=T_{\rm CMB}= 2.73~{\rm K}$.

 The primordial mesons before
 their decays polarize the Dirac fermion vacuum
 (as the origin of axial anomaly \cite{riv,ilieva,gip,j})
 and give the
 baryon asymmetry frozen by the CP -- violation.
The
 value of the baryon--antibaryon asymmetry
of the universe following from this axial anomaly was estimated in
\cite{114:a} in terms of the coupling constant of the
superweak-interaction
 \be\label{a1}
 n_b/n_\gamma\sim X_{CP}= 10^{-9}.
 \ee
The boson life-times     $\tau_W=2H_I\eta_W\simeq
\left({2}/{\alpha_W}\right)^{2/3}\simeq 16,~ \tau_Z\sim
2^{2/3}\tau_W\sim 25
 $ determine the present-day visible
baryon density
\be\label{b}\Omega_b\sim\alpha_W=\alpha_{QED}/\sin^2\theta_W\sim0.03.\ee
All these results (\ref{t}) -- (\ref{b})
 testify to that all  visible matter can be a product of
 decays of primordial bosons, and the observational data on CMBR
 can reflect  parameters of the primordial bosons, but not the
 matter at the time of recombination. In particular,
 the length of  the semi-circle on the surface of  the last emission of
photons at the life-time
  of W-bosons
  in terms of the length of an emitter
 (i.e.
 $M^{-1}_W(\eta_L)=(\alpha_W/2)^{1/3}(T_c)^{-1}$) is
 $\pi \cdot 2/\alpha_W$.
 It is close to $l_{min}\sim  210 $ of CMBR,
 whereas $(\bigtriangleup T/T)$ is proportional to the inverse number of
emitters~
 $(\alpha_W)^3 \sim    10^{-5}$.

 The temperature history of the expanding universe
 copied in the ``conformal quantities'' looks like the
 history of evolution of masses of elementary particles in the cold
 universe with the constant conformal temperature $T_c=a(\eta)T=2.73~ {\rm K}$
 of the cosmic microwave background.

\begin{figure}[t]
\vspace{1cm}
 \begin{center}
 \includegraphics[width=0.5\textwidth,clip]{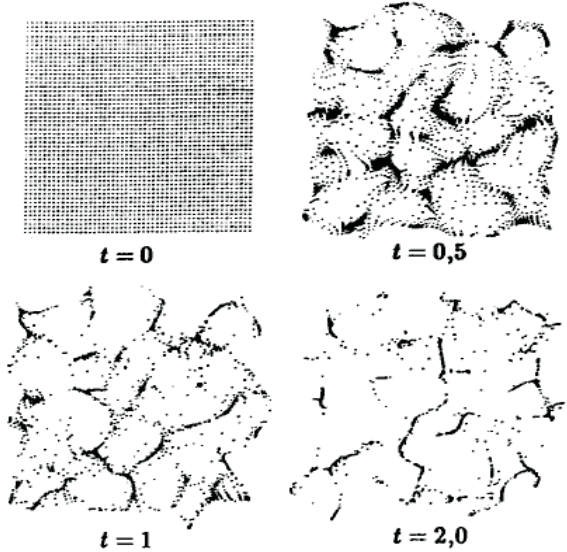}
\caption{
 \label{fig2} {
The
 diffusion of a system of particles moving in the space
 $ds^2=d\eta^2-(dx^i+N^id\eta)^2$ with periodic shift vector $N^i$
 and zero momenta could be understood from analysis of O.D.E.
 $dx^i/d\eta=N^i$ considered in the two-dimensional case in
 \cite{kl}, if we substitute $t= m_{(-)}\eta$ and
 $N^i \sim \frac{x^i}{r}\sin m_{(-)}r$ in the equations,
 where $ m_{(-)}$ is defined by Eq.~(\ref{2-22}).
  }}
\end{center}
\end{figure}

\subsection{Large-scale structure of the matter distribution}
 In the contrast to standard cosmological perturbation theory
 \cite{Lif, kodama, bard}
   the diffeo-invariant version of the perturbation theory
 do not contain time derivatives that are responsible for
the CMB ``primordial power spectrum'' in the inflationary model
\cite{bard}. However, the diffeo-invariant version of the Dirac
Hamiltonian approach to GR gives us another possibility to explain
the CMBR ``spectrum'' and other topical problems of cosmology by
cosmological creation of the vector bosons considered above. The
equations describing the longitudinal vector bosons
 in SM, in this case, are close to
 the equations that follow from
 the Lifshits perturbation theory and are  used, in  the inflationary model, for
 description of the ``power primordial spectrum'' of the CMB radiation.

 The next differences are a nonzero shift vector and  spatial oscillations of
 the scalar potentials determined by $\hat m^2_{(-)}$ (see Fig. \ref{fig2}).
 In the scale-invariant version of cosmology \cite{039}, the
  SN data dominance of stiff state $\Omega_{\rm Stiff}\sim 1$ determines the parameter
  of spatial oscillations
  \mbox{$\hat m^2_{(-)}=\dfrac{6}{7}H_0^2[\Omega_{\rm R}(z+1)^2+\dfrac{9}{2}\Omega_{\rm
  Mass}(z+1)]$}. The redshifts  in the recombination
  epoch $z_r\sim 1100$ and the clustering parameter \cite{kl}
 $
 r_{\rm clustering}=\dfrac{\pi}{\hat m_{(-)} }\sim \dfrac{\pi}{
 H_0\Omega_R^{1/2} (1+z_r)} \sim 130\, {\rm Mpc}
 $
  recently
 discovered in the researches of a large scale periodicity in redshift
 distribution \cite{a1,a2}
 lead to a reasonable value of the radiation-type density
  $10^{-4}<\Omega_R\sim 3\cdot 10^{-3}<5\cdot 10^{-2}$ at the time of this
  epoch.

%%%%%%%%%%%%%%%%%%%%%%%%%%%%%%%%%%%%%%%%%%%%%%%%%%%%%%
%%%%%%%%%%%%%%%%%%%%%%%%%%%

\section{Conclusions}

 We supposed that the Universe was created in a specific reference frame, where the
 Hamiltonian approach to GR is constructed in the finite
 space-time with the diffeomorphisms keeping the frame of
 reference. This frame is remembered by the products of decay
 of the primordial massive vector bosons  created from the Bogoliubov
 stable vacuum.

 The physical content of the Universe is described by both the
 relativistic invariants (of the type of amplitudes of scattering) and
 relativistic covariant quantities like diffeo-invariant time,
 finite volume, temperature, density, etc. Therefore, in
 contrast to the S-matrix approach that depends on only
 relativistic invariants \cite{19}, for the complete description
 of the Universe  we need a set of diffeo-invariant physical
 quantities changing under the Lorentz
 relativistic transformations
 of the type of the dipole component of the CMBR appearing in
 the frame of an Earth observer. Therefore, the quantum creation
 of the Universe in the finite space-time requires the separation
 of the transformations of frames from the diffeomorphisms in the
 context of the David Hilbert formulation of GR \cite{H}.

  Moreover, just this separation (including the choice
   of an evolution parameter in GR as a cosmological scale factor)
  simplifies the Hamiltonian
  equations and
   leads to exact resolution of the energy constraint
 with respect to  the canonical momentum of the scale factor.
 The positive and negative
 values
 of this momentum in
 the ``field space of events'' play the role of the
 generators of evolution  forward and backward, respectively.
 These values of the momentum   onto equations of motion can be called the ``frame energies''.

  The solution of the problem of the ``negative frame energy''  by the
 primary quantization and the
 secondary one (on the analogy of the pathway passed by QFT in
 the 20th century) reveals in GR all attributes of
 the theory of superfluid quantum liquid:
  Landau-type absence of ``friction'',  London-type WDW  wave function,
   and  Bogoliubov-type condensate of quantum universes.

 The postulate of the  quantum Bogoliubov vacuum as the state with
 the minimal ``energy'' leads to
   the absolute beginning of geometric time.
 The fundamental principle of the Hermitian
 Hamiltonian of the evolution in the field space of events
 keeps only  the  potential perturbations of the scalar metric
 components in  contrast to the standard cosmological
 perturbation theory  \cite{Lif}  keeping only the kinetic
 perturbations and the ``friction'' term in the action that is
 responsible for the ``primordial power spectrum'' in
 the inflationary model \cite{bard}.
However, the Quantum Gravity considered as the theory of
superfluidity gives  us
 a  possibility  to explain this ``spectrum'' and
  other topical problems of cosmology
  by  the cosmological creation of the primordial W-, Z- bosons
  from vacuum, when
 their Compton length coincides with the universe horizon.

  The Einstein correspondence
   principle identifies the conformal quantities with the ``measurable''  ones,
   and the uncertainty  principle  establishes the point of
   the beginning of the cosmological creation of the primordial W-, Z- bosons
  from vacuum due to their mass singularity at the moment
  $a_I^2\simeq 10^{-29},H_I^{-1}\simeq 1$ mm.
 The equations describing the longitudinal
 vector bosons
 in SM, in this case, are close to the equations
 of the inflationary model used for
 description of the ``power primordial spectrum'' of the CMB radiation.
 We listed the set of theoretical and observational
 arguments in favor of that the CMB radiation can be
 a final product of  primordial vector W-, Z- bosons cosmologically created
 from the vacuum.

\vspace{1cm}

{\bf Acknowledgement}

\medskip

The authors are grateful to  D.B. Blaschke,  A.A. Gusev, A.V.
Efremov,
 E.A. Kuraev,   V.V. Nesterenko, V.B. Priezzhev,  and S.I. Vinitsky
 for interesting and critical
discussions. AFZ is grateful to the National Natural Science
Foundation of China (NNSFC) (Grant \# 10233050)  for a partial
financial support.

\section*{Appendix A: The energy density in the massive electrodynamics}

\renewcommand{\theequation}{A.\arabic{equation}}

\setcounter{equation}{0}

As the model of the matter let us consider massive electrodynamics
in GR
 \be \label{1-1}
 S=\int d^4x\sqrt{-g}\left[-\frac{\vh_0^2}{6}R(g)
 +{\cal L}_{\rm m}\right],
 \ee
 where ${\cal L}_{\rm m}$ is the Lagrangian of the massive vector and spinor
 fields
\be \label{1-2}
 {\cal L}_{\rm
 m}=-\frac{1}{4}F_{\mu\nu}
 F_{\alpha\beta}g^{\mu\alpha}g^{\nu\beta}-M_0^2 A_\mu A_\nu
 g^{\mu\nu}-\widetilde{\psi}i\gamma^\sigma
(D_{\sigma}-ieA_\sigma)\Psi-m_0\widetilde{\psi}\hat \Psi
 \ee
 $F_{\mu\nu}=\partial_\mu A_\nu-\partial_\nu A_\mu$ is the stress tensor,
 \be\label{1-3}
 D_{\delta}=\partial_{\delta}-i\frac{1}{2}[\gamma_{(\alpha)}
 \gamma_{(\beta)}]\sigma_{\delta(\alpha)(\beta)},
 \ee
is the Fock covariant derivative \cite{fock29},
 $\gamma_{(\beta)}=\gamma^\mu e_{(\beta)\mu}$ are the Dirac $\gamma$-matrices,
 summed with tetrads $e_{(\beta)\nu}$, and
 $\sigma_{\sigma(\alpha)(\beta)}=e^\nu_{(\beta)}(\nabla_\mu
 e_{(\alpha)\nu})$ are
  coefficients of spin-connection \cite{fock29,ll}.

 The Lagrangian of the massive fields (\ref{1-2})  can be
 rewritten
 in terms of the Lichnerowicz variables
 \be\label{1-21}
 {A^{L}}_{\mu}={A}_{\mu},~~~~~~~~~~ \Psi^{L}=a^{3/2}\psi^{3}\Psi,~
 \ee
 that lead to fields with masses depending on the scale factor $a\psi^{2}$
 \be\label{1-22}
  m_{(\rm L)}=m_0a\psi^{2}=m\psi^{2},~~~~~M_{(\rm
  L)}=M_0a\psi^{2}=M\psi^{2}.
  \ee
  These fields are in the space defined by the component of the frame
 \bea \label{1-23}
 \omega^{(\rm L)}_{(0)}&=&\widetilde{\psi}^4~\widetilde{N}_ddx^0,\\\label{1-24}
 \omega^{(\rm L)}_{(a)}&=&{\bf e}_{(a)i}(dx^i+N^i dx^0).
 \eea
  with the unit metric determinant $|{\bf e}|=1$.

  As the result, the Lagrangian of the matter fields (\ref{1-2}) takes the form
  \bea\nonumber %\label{00Lm}
  \sqrt{-g}{\cal L}_{\rm m}(A,\widetilde{\psi},\Psi)=
 \frac{1}{i}\widetilde{\psi}^L\gamma^{0}\left(\partial_0-
 N^k\partial_k+\frac{1}{2}\partial_lN^l-ie A_0\right)\Psi^L -\widetilde{N}_d{\cal H}_\Psi+\\
 \label{1-25}+\widetilde{N}_d\left[-J_{5(c)}v_{[ab]}\varepsilon_{(c)(a)(b)}
 +\frac{\widetilde{\psi}^4}{2}\left(v_{i(\rm A)}v^i_{(\rm A)}-
 \frac{1}{2}F_{ij}F^{ij}\right)-\widetilde{\psi}^8{M^2}A^2_{(b)}-
 \frac{\pi_0^2}{M^2}\right]-\\\nonumber
 +\widetilde{N}_d~\widetilde{\psi}^6 m\widetilde{\psi}^L\Psi^L-\pi_0[{N^iA_i-A_0}]~,
 \eea
 where the Legendre transformation
 $A_0^2/(2\widetilde{N}_d)=\pi_0A_0-\widetilde{N}_d\pi^2_0/2$ with the subsiduary field $\pi_0$
 is used for linearizing the massive term;
  \be\label{1-26} {\cal
 H}_\Psi=
 -\widetilde{\psi}^4[i\widetilde{\psi}^L\gamma_{(b)}D_{(b)}\Psi^L +J^0_5 \sigma-
 \partial_kJ^k]
 \ee
 is the Hamiltonian density of the fermions,
 \bea\label{1-27}
 v_{[ab]}&=&\frac{1}{2}\left({\bf
e}_{(a)i}v^i_{(b)}-{\bf
 e}_{(b)i}v^i_{(a)}\right),\\\label{1-28}
 D_{(b)}\Psi^L&=&[\partial_{(b)}-\frac{1}{2}
 \partial_k {\bf e}^k_{(b)}-ieA_{(b)}]\Psi^L,\\\label{1-29}
 v_{i(\rm A)}&=&\frac{1}{\psi^4\widetilde{N}_d}[\partial_0A_i-\partial_iA_0+F_{ij}N^j]
 \eea
 are the field  velocities, and
 \be\label{1-30}
 J_{5(c)}=\frac{i}{2}(\widetilde{\psi}^L\gamma_5
 \gamma_{(c)}\Psi^L)~,~~~~~~
 J^0_5=\frac{i}{2}(\bar\Psi^L\gamma_5\gamma^0\Psi^L),~~~~~~~~
 J_{k}=\frac{i}{2} \bar\Psi^L\gamma_{k}\Psi^L
 ~;
 \ee
 are the currents, $\sigma=\sigma_{(a)(b)|(c)}
 \varepsilon_{(a)(b)(c)}$, where $\varepsilon_{(a)(b)(c)}$ denotes the Levi-Civita tensor.

 The  canonical conjugated momenta take the  form
\bea\label{1-31}
 P_\vh&=&-2V_0\frac{\partial_0\vh}{N_0}~~~~~~~~~
 ~~~~~=-2V_0\frac{d\vh}{d\zeta}\equiv-2V_0
 \vh'\\\label{1-32}\overline{p_{\psi}}&=&\frac{\partial {\bf K}[\vh|e]}{\partial
 (\partial_0\ln{{\widetilde{\psi}}})}~~~~~~~~~~~~~=-8\vh^2{\overline{v}},
 \\\label{1-33}
 p^i_{(b)}&=&\frac{\partial [{\bf K}[\vh|e]+\sqrt{-g}{\cal L}_{\rm
 m}]}{\partial(\partial_0{\bf e}_{(a)i})}
 ={\bf e}^i_{(a)}\left[\frac{\vh^2}{3} v_{(a b)}-J_{5(c)}
 \varepsilon_{(c)(a)(b)}\right],\\\label{1-34}
 P^i_{\rm (A)}&=&\frac{\partial [\sqrt{-g}{\cal L}_{\rm
 m}]}{\partial(\partial_0{ A}_{i})}~~~~~~~~~~~~=\widetilde{\psi}^4v^i_{\rm
 (A)},\\\label{1-35}
 P_{\rm (\Psi)}&=&\frac{\partial [\sqrt{-g}{\cal L}_{\rm m}]}
 {{\partial(\partial_0{\Psi^L })}}
 ~~~~~~~~~~~~=
 \frac{1}{i}\widetilde{\psi}^L\gamma^{0}.
 \eea
 Then, the action  (\ref{1-1}) one can be represented in the Hamiltonian form
 \be\label{1-36}
 S=\int dx^0\left[-P_{\vh}\partial_0\vh+
 N_0\frac{P^2_\vh}{4V_0}+\int d^3x
 \left(\sum\limits_{{F}
 } P_{F}\partial_0F
 +{\cal C}-\widetilde{N}_d T_{0t}^0\right)\right],
 \ee
 where $P_{F}$ is a set of the field momenta (\ref{1-32}) -- (\ref{1-35}),
\be\label{1-37}
 T_{0t}^0= \widetilde{\psi}^{7}\hat \triangle
\widetilde{\psi}+
  \sum\limits_{I=0,4,6,8} \widetilde{\psi}^I\tau_I,
 \ee
 is the sum of the Hamiltonian densities including the gravity density
  \bea\label{h31}
 \widetilde{\psi}^{7}\hat \triangle
 \widetilde{\psi}&\equiv&\widetilde{\psi}^{7}\dfrac{4\varphi^2}{3}
 \partial_{(b)}\partial_{(b)}\widetilde{\psi},\\\label{h32}
 \tau_{I=0}&=&\dfrac{6\widetilde{p}_{(ab)}\widetilde{p}_{(ab)}}{\vh^2}
 -\dfrac{16}{\vh^2}\overline{p_{\psi}}^2+\frac{\pi_0^2}{2a^2M^2},\\\label{h33}
 \tau_{I=4}&=&\frac{P_{i(\rm A)}P^i_{(\rm A)}+
 F_{ij}F^{ij}}{2}-[i\widetilde{\psi}^L\gamma_{(b)}D_{(b)}\Psi^L +J^0_5 \sigma-
 \partial_kJ^k],\\\label{h34}
 \tau_{I=6}&=&m \widetilde{\psi}^L\Psi^L,\\\label{h35}
 \tau_{I=8}&=&\dfrac{\varphi^2}
  {6}R^{(3)}({\bf e})+\frac{M^2A^2_{(b)}}{2},
\eea
  here $\widetilde{p}_{(ab)}=\frac{1}{2}({\bf e}^i_{(a)}\widetilde{p}_{(b)i}+
  {\bf e}^i_{(b)}\widetilde{p}_{(a)i})$, $\widetilde{p}_{(b)i}={p}_{(b)i}+{\bf e}^i_{(a)}
  \varepsilon_{(c)(a)(b)}J_{(c)}$,
 \be\label{1-41}
 {\cal C}=A_0[\partial_iP^i_{(A)}+eJ_0-\pi_0]+N_{(b)}
  {T}^0_{(b)t} +\lambda_0\overline{p_\psi}+ \lambda_{(a)}\partial_k{\bf e}^k_{(a)}
 \ee
 denotes the sum of the constraints, where
 $J_0=\widetilde{\psi}^L\gamma_0\Psi^L$ is the zero component of the current; $A_0, N_d, N^i,
 \lambda_0,
\lambda_{(a)}$
 are the Lagrange multipliers  including the Dirac condition
 of the minimal 3-dimensional hyper-surface \cite{dir}
\be\label{1-42}
 p_{\widetilde{\psi}}=\overline{v}=0 \to
 (\partial_0-N^l\partial_l)\log{
 \widetilde{\psi}}=\frac16\partial_lN^l,
 \ee%{proi1}
 that gives a positive value of the Hamiltonian density
 (\ref{h32}), and
 \bea\label{1-43}
 {T^0_{(a)}}_t= &-&\overline{p_{\psi}}\partial_{(a)}
 \widetilde{\psi}+\frac{1}{6}\partial_{(a)}
 (\overline{p_{\psi}}{\overline\psi}) +
 2p_{(b)(c)}\gamma_{(b)|(a)(c)}-\partial_{(b)}p_{(b)(a)}+
 \\\nonumber%\label{1-43}
%{T^0_{(a)}}_{({\rm m})}&=&
&-& \frac{1}{i}\bar\Psi^I\gamma^{0}\partial_{(a)}
\Psi^I-\frac{1}{2i}
\partial_{(a)}\left(\bar\Psi^I\gamma^{0} \Psi^I\right) -P^i_{(\rm
A)}F_{ik}{\bf e}^k_{(a)}-\pi_0A_{(a)}. \eea are the components of
the total energy-momentum tensor ${T}^0_{(a)}={T}^0_{i}{\bf
e}^i_{(a)}$.

\section*{Appendix B: Diffeo-Invariant Content of a Relativistic String}

\renewcommand{\theequation}{B.\arabic{equation}}

\setcounter{equation}{0}

To illustrate the invariant Hamiltonian reduction by putting
strong constraints like (\ref{non1}) let us consider the action
for a relativistic string \cite{1} in the form which has been done
by L. Brink, P. Di Vecchia, P. Howe \cite{2}

\be\label{ss} S=-\frac{\gamma}{2}\int\int d^2
u\sqrt{-g}g^{\alpha\beta}\partial_{\alpha}x^{\mu}
(u_0,u_1)\partial_{\beta}x_{\mu}(u_0,u_1),~~~~~~ (u_0,u_1)=(\tau,
\sigma),\ee where $x^{\mu}(\tau,\sigma)$ -- string coordinates
given in d-dimension space-time $(\mu=0,1,2,\ldots ,d-1)$,
$g_{\alpha\beta}(u_0,u_1)$ -- is a second-rank metric tensor on
the string surface (two dimensional Riemannian space $u_0,u_1$).

Now let us consider the Hamiltonian scheme which is based on the
Arnowitt--Deser--Misner parametrization of metric tensor
$g_{\alpha\beta}$ \cite{3} \be\label{mm} g_{\alpha\beta}=\Omega^2
\begin{pmatrix}
N_0^2-N_1^2& N_1\\
N_1 & -1
\end{pmatrix},~~~ g^{\alpha\beta}=\frac1{\Omega^2N_0^2}\begin{pmatrix}
1&N_1\\
N_1& N_1^2-N_0^2
\end{pmatrix},~~~ \sqrt{-g}=\Omega^2N_0
\ee with the conformal invariant interval \be\label{me}
ds^2=g_{\alpha\beta}du^{\alpha}du^{\beta}=\Omega^2[N_0^2d\tau^2-
(d\sigma+N_1d\tau)^2], \ee where $N_0$ and $N_1(\tau,\sigma)$ are
known in GR as the lapse function and ``shift vector'',
respectively (compare formulas (\ref{1adm})).

The action (\ref{ss}) after the substitution (\ref{mm}) does not
depend on the conformal factor $\Omega$ and takes the form
\be\label{a4} S=-\frac{\gamma}2
\int_{\tau_1}^{\tau_2}d\tau\int^l_0d\sigma
\left[\frac{(\dot{x}_{\mu}-N_1x'_{\mu})^2}{N_0}-N_0x'{}^2\right],
\ee where $\dot{x}_{\mu}=\partial_{\tau}x_{\mu}$,
$x'_{\mu}=\partial_{\sigma}x_{\mu}$ and
$\dot{x}_{\mu}-N_1x'_{\mu}={D}{}x_{\mu}$ is the covariant derivative
with respect to the metric (\ref{me}). The action (\ref{a4}), the
metric (\ref{me}) and the covariant derivative $Dx_{\mu}$ are
invariant under the ``kinemetric'' transformation
$\tau\rightarrow\tilde{\tau}=f_1(\tau),~
\sigma\rightarrow\tilde{\sigma}=f_2(\tau,\sigma)$ that are similar
to group of transformation in GR (\ref{zel}), (\ref{kine}). The
``kinemetric'' transformations of the differentials
$d\tilde{\tau}=\dot{f}_1(\tau)d\tau$,
$d\tilde{\sigma}=\dot{f}_2(\tau,\sigma)d\tau+{f}_2{}'(\tau,\sigma)d\sigma$
correspond to transformations of the string coordinates (compare
with (\ref{kine}) in the text)
\begin{align}
x_{\mu}{}'(\tau,\sigma)=&~\tilde{x}_{\mu}{}'(\tilde{\tau},\tilde{\sigma})f_2{}
'(\tau,\sigma),\notag\\\label{kintr}
\dot{x}_{\mu}(\tau,\sigma)=&~\dot{\tilde{x}}_{\mu}(\tilde{\tau},\tilde{\sigma})
\dot{f}_1(\tau)+
\tilde{x}_{\mu}{}'(\tilde{\tau},\tilde{\sigma})\dot{f}_2(\tau,\sigma),\\
N_0(\tau,\sigma)=&~\tilde{N}_0(\tilde{\tau},\tilde{\sigma})
\frac{\dot{f}_1(\tau)}{f_2{}'(\tau,\sigma)},~~~
N_1(\tau,\sigma)=\tilde{N}_1(\tilde{\tau},\tilde{\sigma})
\frac{\dot{f}_1}{\tilde{f}_2{}'}+\frac{\dot{f}_2}{\tilde{f}_2{}'},\notag
\end{align}
\be D_{\tau}x_{\mu}(\tau,\sigma)=\dot{f}_1(\tau)D_{\tilde{\tau}}
\tilde{x}_{\mu}(\tilde{\tau},\tilde{\sigma}) \ee

The variation $S$ with respect to $N_0$ and $N_1$ leads to
equation for determination $N_0, N_1$
\begin{align}
\frac{\delta S}{\delta
N_0}=\frac{(Dx_{\mu})^2}{N_0^2}+x'^2=0~\Rightarrow &~
N_0^2=\frac{(\dot{x}x')^2-\dot{x}^2x'^2}{(x'^2)^2},\\
\frac{\delta S}{\delta
N_1}=\frac{(x'{}_{\mu}Dx^{\mu})}{N_0}~\Rightarrow &~
N_1=\frac{(\dot{x}x')}{x'^2},
\end{align}
The substitution of these equation in action (\ref{a4}) converts it
into the standard Nambu--Goto action of the relativistic string
\cite{1}
\be S=-\gamma\int_{\tau_1}^{\tau_2}d\tau\int^l_0
d\sigma\sqrt{(\dot{x}x')^2-\dot{x}^2x'^2}.\ee

One can construct the Hamiltonian form of $S$. The conjugate momenta
are determined by

\be p_{\mu}=\frac{\delta S}{\delta
\dot{x}^{\mu}}=\gamma\frac{\dot{x}_{\mu}-N_1x_{\mu}{}'}{N_0}\Rightarrow
\dot{x}_{\mu}=N_0\frac{p_{\mu}}{\gamma}+N_1x_{\mu}{}' \ee and
density of Hamiltonian is obtained by the Legendre transformation
\be\label{ham}
H=p_{\mu}\dot{x}^{\mu}-L=N_0\frac{P^2+\gamma^2x'^2}{2\gamma}+
N_1(px'), \ee then \be S=-\int_{\tau_1}^{\tau_2} d\tau\int_0^l
d\sigma\left[p_{\mu}\dot{x}^{\mu}-H\right]. \ee The secondary class
constraints arises by varying $S$ with respect to $N_0, N_1$
\be\label{con} \frac{\delta S}{\delta
N_0}=\frac{p^2+\gamma^2x'^2}{2\gamma}=0,~~ \frac{\delta S}{\delta
N_1}=(px')=0. \ee The equations of motion take the form
\begin{align}
\frac{\delta S}{\delta
x^{\mu}}=&~\dot{p}_{\mu}-\frac{\partial}{\partial\sigma}(\gamma
N_0x_{\mu}'+N_1p_{\mu})=0\notag,\\
&\hphantom{\frac{\delta S}{\delta
x^{\mu}}=~\dot{p}_{\mu}-\frac{\partial}{\partial\sigma}(\gamma
N_0x_{\mu}'+N_1p_{\mu})=0}\label{ph}\\
 \frac{\delta S}{\delta
p^{\mu}}=&~\dot{x}_{\mu}-N_0\frac{p_{\mu}}{\gamma}+N_1x_{\mu}'=0.\notag
\end{align}

The standard gauge-fixing method is to fix the second class
constraints (orthonormal gauge) $N_0=1, N_1=0.$ In this case the
equations of motion (\ref{ph}) reduce to d'Alambert  ones for
$x_{\mu}$ \be \dot{p}_{\mu}=\gamma x_{\mu}{}'',~~
\dot{x}_{\mu}=\frac{p_{\mu}}{\gamma}\Rightarrow
\ddot{x}_{\mu}-x_{\mu}{}''=0, \ee the conformal interval (\ref{me})
takes the usual form
$$ ds^2=\Omega^2[d\tau^2-d\sigma^2],$$
but the Hamiltonian (\ref{ham}) in view of the constraint
(\ref{con}) is equal to zero $(H=0)$.

There is another way to introduce evolution parameter as the object
reparameterizations in the theory being adequate to the initial
``kinemetric'' invariant system and to construct the non-zero
Hamiltonian. Accordingly to (\ref{13c}) we identify this evolution
parameter with the time-like variable of the ``center of mass'' (CM)
of a string defined as the total coordinate \be
X_{\mu}(\tau)=\frac1{l}\int^l_0 x_{\mu}(\tau,\sigma)d\sigma. \ee
Therefore, the reduction requires to separate the ``center of mass''
variables before variation of the action (\ref{a4}) which after
substitution \be\label{ix}
x_{\mu}=X_{\mu}(\tau)+\xi_{\mu}(\tau,\sigma), ~
x_{\mu}'(\tau,\sigma)=\xi_{\mu}'(\tau,\sigma), \ee \be\label{18}
\int^l_0\xi_{\mu}(\tau,\sigma)=0 \ee takes the form
\begin{multline}
S=\frac{\gamma}{2}\int^{\tau_2}_{\tau_1}d\tau\left\{\dot{X}^2(\tau)\int^l_0
\frac{d\sigma}{N_0(\tau,\sigma)}+2\dot{X}^{\mu}(\tau)\int^l_0
d\sigma \left(\frac{\dot{\xi}_\mu-N_1\xi_{\mu}'}{N_0}\right)+ \right. \\
\left.+\int^l_0
d\sigma\left[\frac{(\dot{\xi}_\mu-N_1\xi_{\mu}')^2}{N_0}-N_0\xi'^2(\tau,\sigma)\right]\right\}.\label{long}
\end{multline}
The usual determination  of the conjugate momenta \be
P_{\mu}(\tau)=\frac{\delta S}{\delta \dot{X}^{\mu}(\tau)}= \gamma
\dot{X}_{\mu}(\tau)\int^l_0\frac{d\sigma}{N_0(\tau,\sigma)}+
\gamma\int^l_0
d\sigma\left(\frac{\dot{\xi}_{\mu}(\tau,\sigma)-N_1\xi_{\mu}'(\tau,\sigma)}{N_0(\tau,\sigma)}\right),
\ee \be \pi_{\mu}(\tau,\sigma)=\frac{\delta
S}{\delta\dot{\xi}^{\mu}(\tau,\sigma)}=\gamma \dot{X}_{\mu}(\tau)
\frac1{N_0(\tau,\sigma)}+\gamma\frac{\dot{\xi}_{\mu}(\tau,\sigma)-N_1\xi_{\mu}'(\tau,\sigma)}
{N_0(\tau,\sigma)}\ee leads to the contradiction because
$P_{\mu}(\tau)$ and $\pi_{\mu}(\tau,\sigma)$ are not independent ,
namely $\int^l_0\pi_{\mu}(\tau,\sigma)d\sigma=P_{\mu}(\tau)$
(compare (\ref{1pi1}), (\ref{2pi1}), (\ref{2pi})).

Thus for the consistent definition  of the momentum of
``center-mass'' $P_{\mu}(\tau)$ and momentum of intrinsic variables
$\pi_{\mu}(\tau,\sigma)$ we have to put strong constraint  in action
(\ref{long}) (compare (\ref{friction})) \be\label{hren} \int
d\sigma\left(\frac{\dot{\xi}_{\mu}(\tau,\sigma)-
N_1\xi_{\mu}'(\tau,\sigma)}{N_0}\right)=0. \ee Then we obtain the
following form of the reduced action: \be S_{\rm
red}=\frac{\gamma}{2}\int d\tau
\left\{\dot{X}^2(\tau)\frac{l}{N(\tau)}+
\int^l_0d\sigma\left[\frac{[\dot{\xi}_{\mu}(\tau,\sigma)-
N_1\xi_{\mu}'(\tau,\sigma)]^2}{N_0(\tau,\sigma)}-
N_0(\tau,\sigma)\xi_{\nu}'{^2}(\tau,\sigma)\right]\right\}, \ee
where $N(\tau)$ is the global lapse function \be\label{glf}
\frac{1}{N(\tau)}=\frac1{l}\int^l_0\frac{d\sigma}{N_0(\tau,\sigma)}=
 \langle N_0^{-1}\rangle.\ee
 For global momenta (compare with (\ref{1pi1})) we get \be
P_{\mu}(\tau)=\frac{\partial S_{\rm
red}}{\partial\dot{X}^{\mu}(\tau)}=\gamma\dot{X}_{\mu}(\tau)\frac{l}{N(\tau)}
\ee and for the local (intrinsic) momenta (compare with
(\ref{2pi1})) \be \pi_{\mu}(\tau,\sigma)=\frac{\partial S_{\rm
red}}{\partial\dot{\xi}^{\mu}(\tau,\sigma)}=
\gamma\frac{\dot{\xi}_{\mu}(\tau,\sigma)-
N_1(\tau,\sigma)\xi_{\mu}'(\tau,\sigma)}{N_0(\tau,\sigma)} \ee with
two strong constraints (\ref{18}) and (\ref{hren}) \be\label{a18}
\int^l_0\xi_{\mu}(\tau,\sigma)d\sigma=0,~~~~
\int^l_0\pi_{\mu}(\tau,\sigma)d\sigma=0. \ee This separation
conserves the group of the ``kinemetric'' transformation
(\ref{kintr}) and leads to Hamiltonian form of reduced action, in
view of \be
\dot{X}_{\mu}(\tau)=\frac{N(\tau)}{l}\frac{P_{\mu}(\tau)}{\gamma};~~~~~~~
\dot{\xi}_{\mu}(\tau,\sigma)=
\frac{N_0(\tau,\sigma)}{\gamma}\pi_{\mu}(\tau,\sigma)+
N_1(\tau,\sigma)\xi_{\mu}'(\tau,\sigma), \ee  we get \be
H=P_{\mu}\dot{X}^{\mu}(\tau)+\pi_{\mu}\dot{\xi}^{\mu}-L=\frac1{2\gamma}
\left\{\frac{N_0(\tau,\sigma)}{l}P^2(\tau)+
N_0(\tau,\sigma)[\pi^2+\gamma^2\xi'{}^2]+2\gamma
N_1(\pi\xi')\right\}, \ee

\be\label{a}
S=\int^{\tau_2}_{\tau_1}d\tau\left\{P_{\mu}(\tau)\dot{X}^{\mu}(\tau)-
N(\tau)\frac{P^2(\tau)}{2\gamma l}+ \int^l_0 d\sigma
\left[(\pi\dot{\xi})-N_0\frac{\pi^2+\gamma^2\xi'{}^2}{2\gamma}-
N_1(\pi\xi')\right]\right\}. \ee The equation of motion is split
into global one \be \frac{\delta
S}{\delta\dot{X}^{\mu}(\tau)}=\dot{P}_{\mu}(\tau)=0,~~~~~
\frac{\delta S}{\delta
P^{\mu}(\tau)}=\dot{X}_{\mu}(\tau)-{N(\tau)}\frac{P_{\mu}}{\gamma
l}=0 \nonumber \ee and local one \be \frac{\delta
S}{\delta\pi^{\mu}(\tau,\sigma)}=\dot{\xi}_{\mu}
(\tau,\sigma)-\frac{N_0}{\gamma}\pi_{\mu}-N_1\xi_{\mu}'=0,~~~
-\frac{\delta S}{\delta\xi^{\mu}(\tau,\sigma)}=\dot{\pi}_{\mu}-
\frac{\partial}{\partial\sigma}(N_1\pi_{\mu}+\gamma
N_0\xi_{\mu}')=0. \ee The variation of the action (\ref{a}) with
respect to $N_0(\tau)$ results in the constraint \be\label{20}
\frac{\delta S}{\delta
N_0(\tau,\sigma)}=\frac{N^2(\tau)}{N^2_0(\tau,\sigma)}\frac{P^2(\tau)}{2\gamma
l^2}+
\frac{\pi^2(\tau,\sigma)+\gamma^2\xi'{}^2(\tau,\sigma)}{2\gamma}=0,
\ee here it is necessary to take into account that the variation
over the global lapse function (\ref{glf}) leads to \be \frac{\delta
N(\tau)}{\delta
N_0(\tau,\sigma)}=\frac{N^2(\tau)}{N_0^2(\tau,\sigma)}. \ee The
variation of the (\ref{a}) with respect to $N_1(\tau,\sigma)$
results in the constraint for local variables \be \frac{\delta
S}{\delta N_1(\tau,\sigma)}=(\pi\xi'(\tau,\sigma))=0. \ee Now we
introduce the Hamiltonian density for local excitations
\be\label{a22}
{\cal{H}}(\tau,\sigma)=-\frac{\pi^2(\tau,\sigma)+\gamma^2\xi'^2(\tau,\sigma)}{2\gamma}
\ee and rewrite (\ref{20}) in the form \be\label{a23}
\frac{N(\tau)}{N_0(\tau,\sigma)l}\sqrt{P^2_{\mu}(\tau)}=
\sqrt{\vphantom{P^2_2}2\gamma{\cal H}(\tau,\sigma)}.\ee One
integrates (\ref{a23}) over $\sigma$ and taking into account the
normalization equality (\ref{glf})  \be\label{e}
\frac1l\int^l_0\frac{N(\tau)}{N(\tau,\sigma)}d\sigma=1, \ee it
 leads to global constraint %(compare (\ref{13ec}))
 \be\label{mst}
M_{\rm st}=\sqrt{P^2_{\mu}(\tau)}=\int^l_0\sqrt{2\gamma{\cal
H}(\tau,\sigma)}d\sigma=l\langle\sqrt{2\gamma{\cal H}}\rangle, \ee
where $M_{\rm st}$ is mass of the string. The local part of the
constraint (\ref{a23}) can be obtained by substitution (\ref{mst})
into (\ref{a23}) \be\label{a26}
\frac{N(\tau)}{N_0(\tau,\sigma)}=\frac{\sqrt{{\cal
H}(\tau,\sigma)}}{\langle\sqrt{{\cal H}}\rangle}, \ee which coincide
with the equation (\ref{13ec}) in the text. Finally, after
substitution (\ref{mst}), (\ref{a26}) into action (\ref{a}) we can
derive the constraint-shell action, whereas the equation
$N(\tau)P^2(\tau)/2\gamma l=\int N_0(\tau,\sigma){\cal
H}(\tau,\sigma)d\sigma$ \be\label{s} S_{\rm
const-shell}=\int^{\tau_2}_{\tau_1}d\tau\{P_{\mu}\dot{X}^{\mu}(\tau)+
\int^l_0d\sigma[\pi_{\mu}(\tau,\sigma)\dot{\xi}^{\mu}(\tau,\sigma)-
N_1(\tau,\sigma)\pi_{\mu}(\tau,\sigma)\xi'_{\mu}(\tau,\sigma)]\}.
\ee Again the variation with respect to $N_1(\tau,\sigma)$ results
in subsidiary condition
\mbox{$(\pi_{\mu}(\tau,\sigma)\xi'{}^{\mu}(\tau,\sigma))=0.$} Now in
the ``center-mass'' coordinate system $P_i(\tau)=0,~~
P_0(\tau)=\int^l_0d\sigma\sqrt{2\gamma{\cal H}(\tau,\sigma)}$ the
action (\ref{s}) takes the form \be\label{a28} S_{\rm const-shell}=
\int^{\tau_2}_{\tau_1}d\tau\int^l_0d\sigma\{\pi_{\mu}(\tau,\sigma)
\dot{\xi}^{\mu}(\tau,\sigma)+ \sqrt{2\gamma{\cal
H}(\tau,\sigma)}\dot{X}_0(\tau)\}, \ee which describes the dynamics
of a local (intrinsic) canonical variables of a string with non-zero
Hamiltonian because (\ref{a28}) can be rewritten in the form
\be\label{a29} S_{\rm
const-shell}=\int^{x_2}_{x_1}dX_0\int^l_0d\sigma\left\{\pi_{\mu}(X_0,\sigma)
\frac{\partial\xi^{\mu}(X_0,\sigma)}{\partial X_0}+
\sqrt{2\gamma{\cal H}(X_0,\sigma)}\right\}, \ee where
$$2\gamma{\cal H}(X_0,\sigma)=-[\pi^2(X_0,\sigma)+
\gamma^2\xi'{}^2(X_0,\sigma)]$$ and time
$dX_0=\dot{X}_0(\tau)d\tau$ is invariant with respect to
$d\tilde{\tau}=f_1d\tau.$

In the gauge-fixing method, by using the kinemetric transformation,
we have to put \mbox{$N_0(\tau,\sigma)=1,$}
\mbox{$N_1(\tau,\sigma)=0$} (this requirement does not contradict to
Eq. (\ref{e}) in view of Eqs. (\ref{mst}), (\ref{a26})). Then
according to \cite{fr}\be\label{a30} \sqrt{{\cal
H}(\tau,\sigma)}=\frac1l\int^l_0d\sigma\sqrt{{\cal
H}(\tau,\sigma)}=\frac{M_{\rm st}}{\sqrt{2\gamma}~l} \ee  means that
the Hamiltonian $\sqrt{2\gamma{\cal H}(\tau,\sigma)}$ is constant.
In this case the equations for the local variables obtained by
varying the action (\ref{a29}) take the form \be \frac{\delta
S}{\delta \pi^{\mu}(X_0,\sigma)}=\frac{\partial
\xi_{\mu}(X_0,\sigma)}{\partial
X_0}-\frac{\pi_{\mu}(X_0,\sigma)}{\sqrt{-(\pi^2+\gamma^2\xi'{}^2)}}=0,
\ee \be \frac{\delta S}{\delta \xi^{\mu}(X_0,\sigma)}=
-\frac{\partial \pi_{\mu}(X_0,\sigma)}{\partial X_0}+\gamma^2
\frac{\partial/\partial\sigma
\xi'_{\mu}(X_0,\sigma)}{\sqrt{-(\pi^2+\gamma^2\xi'{}^2)}}=0. \ee If
we put $\sqrt{-(\pi^2+\gamma^2\xi'{}^2)}=\gamma,~~ (M_{\rm
st}=\gamma l)$, then it  leads again to d'Alambert equation for
$\xi_{\mu}(X_0,\sigma)$ \be \frac{\partial^2
\xi_\mu{(X_0,\sigma)}}{\partial X^2_0}=\frac{\partial^2
\xi_\mu{(X_0,\sigma)}}{\partial \sigma^2}. \ee The general solution
of these equations in class of functions (\ref{a18}) with boundary
conditions for the open string
$\xi'{}_{\mu}(X_0,0)=\xi'{}_{\mu}(X_0,l)=0$ is given as usually by
the Fourier series
\begin{align}
\xi_{\mu}(X_0,\sigma)=&\frac1{2\sqrt{\gamma}}[\Psi_{\mu}(X_0+\sigma)+
\Psi_{\mu}(X_0-\sigma)],\notag\\
\xi'{}_{\mu}(X_0,\sigma)=&\frac1{2\sqrt{\gamma}}[\Psi'{}_{\mu}(X_0+\sigma)-
\Psi'{}_{\mu}(X_0-\sigma)],\label{a47}\\
\pi_{\mu}(X_0,\sigma)=&\gamma\frac{\partial\xi_{\mu}(X_0,\sigma)}{\partial
X_0}=\frac{\sqrt{\gamma}}{2}[\Psi'{}_{\mu}(X_0+\sigma)+
\Psi'{}_{\mu}(X_0-\sigma)],\notag
\end{align}
where \be\label{a32} \Psi_{\mu}(z)=i\sum_{n\neq
0}\frac{\alpha_{n\mu}}{n}e^{-\frac{i\pi n}{l}z},~~
\Psi'{}_{\mu}(z)=\frac{\pi}l\sum_{n\neq
0}\alpha_{n\mu}e^{-\frac{i\pi n}{l}z} \ee (it does not contain zero
harmonics $(n\neq 0)$). The substitution of $\xi_{\mu}$ and
$\pi_{\mu}$ in this form into (\ref{a22}) and taken into
consideration (\ref{a30}) leads to density of Hamiltonian
\be\label{a33} {\cal H}=-\frac14[\Psi'{}^2_{\mu}(X_0+\sigma)+
\Psi'{}^2_{\mu}(X_0-\sigma)]=\frac{M^2_{\rm st}}{2\gamma l^2}. \ee
For the constraint $(\pi\xi')=0$ in terms of the vector
$\Psi'{}_{\mu}$ (\ref{a47}) we obtain \be\label{a34}
\pi_{\mu}(X_0,\sigma)\xi'{}^{\mu}(X_0,\sigma)=
\frac14[\Psi'{}_{\mu}^2(X_0+\sigma)-\Psi'{}_{\mu}^2(X_0-\sigma)]=0,
\ee then from (\ref{a33}) and (\ref{a34}) finally we get
\be\label{eq}
\Psi_{\mu}'{}^2(X_0+\sigma)=\Psi_{\mu}'{}^2(X_0-\sigma)=-\frac{M_{\rm
st}^2}{l^2\gamma}.\ee It means that $\Psi_{\mu}'(z)$ is the
modulo-constraint space-like vector and in terms of its
representation (\ref{a32}) the equalities (\ref{eq}) can be
rewritten \be\label{a36}
-\Psi_{\mu}'{}^2(z)=\frac{\pi^2}{l^2}\sum^{\infty}_{k=-\infty}
L_ke^{-i\frac{\pi k}{l}z}=\frac{M^2_{\rm st}}{l^2 \gamma},\ee where
$$L_k=-\sum_{n\neq k,0}\alpha_{n\mu}\alpha^{\mu}_{k-n},~~~~
L_k^{*}=L_{-k}~.$$ Now we can see that the zero harmonic of this
constraint determines the mass of a string \be M_{\rm
st}^2=\pi^2\gamma L_0=-\pi\gamma
\sum_{n\neq0}\alpha_{n\mu}\alpha^{\mu}_{-k},~~~~
\alpha^{\mu}_{-k}=\alpha^{*}_k{}^{\mu} \ee and coincides with
standard definition in string theory \cite{1}, however the non-zero
harmonics of constraint (\ref{a36}) \be\label{a37}
L_{k\neq0}=-\sum_{n\neq0,k}\alpha_{nk}\alpha^{\mu}_{k-n}=0 \ee
strongly differ from the standard theory because they do not depend
on the global motion(do not depend on $P_{\mu}$) and do not contain
the interference term $P_{\mu}\Psi^{\mu}{}'$, because of our
constraint (\ref{eq}) we can rewrite \be
l^2\gamma\Psi_{\mu}'{}^2(z)+P^2_{\mu}=0 \ee instead standard one
\cite{1} \be (P_{\mu}+l\sqrt{\gamma}\Psi_{\mu}'(z))^2=0. \ee
Therefore our approach coincides with the R{\"o}hrlich one \cite{4},
which is based on the gauge condition $P_{\mu}\xi^{\mu}=0,~~~
P_{\mu}\pi^{\mu}=0 \Rightarrow P_{\mu}\alpha_n^{\mu}=0,~~~ n\neq0$.
One use of that condition for eliminating the time components
$\xi_0, \pi_0$ being constructed in the frame of reference $(P_i=0)$
leads to formula (\ref{a47})--(\ref{a37}), where $\Psi'_{0}$ and
$\alpha_{n0}$ are equal to zero.

\end{document}